\documentclass[useAMS,usenatbib]{mn2e}
\usepackage{epsfig,amsmath,natbib}

\usepackage{graphicx}
\usepackage{txfonts}
\usepackage{ctable}
\usepackage{threeparttable}
\usepackage{subfigure}
\def\be{\begin{equation}}
\def\ee{\end{equation}}

\def\la{\lsim}
\def\ga{\gsim}

\def\Msun{M_{\odot}}
\def\Zsun{Z_{\odot}}
\def\Lsun{L_{\odot}}

\def\E3{{\cal E}_{\rm g}^{III}}

\title[Pop III star formation with most iron-poor stars]{Limits on Pop III star formation with the most iron-poor stars}
\author[de Bennassuti et al.]{M. de Bennassuti$^{1, 2}$\thanks{E-mail:
matteo.debennassuti@oa-roma.inaf.it}, S. Salvadori$^{3,4,5}$, 
R. Schneider$^{2,5}$, R.Valiante$^{2}$, K. Omukai$^{6,5}$\\
$^{1}$Dipartimento di Fisica, Sapienza, Universit$\grave{a}$ di Roma, Piazzale Aldo Moro 5, 00185, Roma, Italy\\
$^{2}$INAF - Osservatorio Astronomico di Roma, Via di Frascati 33, 00078 Monte Porzio Catone, Italy\\
$^{3}$Laboratoire d'Etudes des Galaxies, Etoiles, Physique et Instrumentation GEPI, Observatoire de Paris, Place Jules Jannsen, 92195 Meudon, Paris, France\\ 
$^{4}$Kapteyn Astronomical Institute, University of Groningen, Landleven 12, 9747 AD Groningen, The Netherlands\\
$^{5}$Kavli Institute for Theoretical Physics, Kohn Hall, University of California, Santa Barbara, CA 93106\\
$^{6}$Astronomical Institute, Tohoku University, Aoba, Sendai 980-8578, Japan}
\begin{document}

\label{firstpage}

\date{Accepted . Received}

\pagerange{\pageref{firstpage}--\pageref{lastpage}} 

\pubyear{}

\maketitle

\begin{abstract}
We study the impact of star-forming mini-haloes, and the Initial Mass Function 
(IMF) of Population III (Pop~III) stars, on the Galactic halo Metallicity 
Distribution Function (MDF) and on the properties of C-enhanced and C-normal 
stars at [Fe/H]$<-3$. For our investigation we use a data-constrained merger 
tree model for the Milky Way formation, which has been improved to self-consistently 
describe the physical processes regulating star-formation in mini-haloes, 
including the poor sampling of the Pop~III IMF. We find that only when 
star-forming mini-haloes are included the low-Fe tail of the MDF is 
correctly reproduced, showing a plateau that is built up by C-enhanced 
metal-poor (CEMP) stars imprinted by primordial faint supernovae. The 
incomplete sampling of the Pop~III IMF in inefficiently star-forming 
mini-haloes ($<10^{-3}\Msun$~yr$^{-1}$) strongly limits the formation 
of Pair Instability Supernovae (PISNe), with progenitor masses 
$m_{\rm popIII} = [140 - 260] \,\Msun$, even when a flat Pop~III IMF 
is assumed. Second-generation stars formed in environments polluted 
at $>50\%$ level by PISNe are thus extremely rare, corresponding to 
$\approx 0.25\%$ of the total stellar population at [Fe/H]$<-2$, which 
is consistent with recent observations. The low-Fe tail of the MDF 
strongly depends on the Pop III IMF shape and mass range. Given the 
current statistics, we find that a flat Pop~III IMF model with 
$m_{\rm popIII} = [10 - 300]\, \Msun$ is disfavoured by observations. 
We present testable predictions for Pop III stars extending down to lower masses,
with $m_{\rm popIII} = [0.1-300]\, \Msun$.
\end{abstract}

%%%%%%%%%%%%%%%%%%%%%%%%%%%%%%%%%%%%%%%%%%%%%%%%%%%%%%%%%%%%%%%%%%%%%%%%%%%
\section{Introduction}
\label{sec:intro}
%%%%%%%%%%%%%%%%%%%%%%%%%%%%%%%%%%%%%%%%%%%%%%%%%%%%%%%%%%%%%%%%%%%%%%%%%%
According to the standard $\Lambda$ Cold Dark Matter ($\Lambda$CDM) 
model for structure formation, the first stars formed at $z\approx 20$ 
in low-mass dark matter ``mini-haloes'', with total masses $M\approx 
[10^6-10^7]\,\Msun$ and virial temperatures $T_{\rm vir} < 10^4$~K \citep[e.g.
][for a recent review]{abel02,bromm13}. At these low temperatures, and 
in gas of primordial composition, the only available coolant is molecular 
hydrogen, $H_2$, which can be easily photo-dissociated by Lyman Werner 
photons (LW, $E=11.2-13.6$~eV) produced by the first and subsequent 
stellar generations \citep[e.g.][]{haiman97,omukai99,machacek01}. The ability of mini-haloes 
to efficiently form stars is therefore highly debated, and it critically 
depends upon the specific physical properties of these low-mass systems, 
such as their formation redshift, gas temperature, and gas metallicity, 
which determine the cooling efficiency of the gas (e.g. Omukai 2012).

During the epoch of reionization, furthermore, the gas surrounding 
star-forming galaxies gets ionized and heated up to temperatures 
$T > 10^4$~K (e.g. Maselli et al. 2003; Graziani et al. 2015). 
Gas infall is thus suppressed 
in low temperature mini-haloes born in ionized cosmic regions and it 
is completely quenched after the end of reionization, at $z\approx 6$ 
\citep[e.g.][]{gnedin00,okamoto08,noh14,graziani15}. Because of the intrinsic 
fragility of mini-haloes, many galaxy formation models neglect 
star-formation in these low-mass systems (e.g. Bullock et al. 2015). 
However, mini-haloes likely played an important role in the early 
Universe, being the nursery of the first stars and the dominant halo 
population, which likely regulated the initial phases of reionization 
and chemical-enrichment \citep[e.g.][]{salvadori14,wise14}. 

During last years, observational evidences of the importance 
of mini-haloes have been provided by ultra-faint dwarf galaxies, 
the faintest and most metal-poor galaxy population in the Local 
Group \citep{simon07}. These galaxies, which have total luminosities 
$L<10^5\Lsun$, have been proposed to be the living fossil of the 
mini-haloes which overcame radiative feedback processes and managed 
to form stars before the end of reionization \citep{salvadori09,
bovill09,munoz09}. Deep color-magnitude diagrams of ultra-faint dwarfs 
have confirmed these theoretical predictions, showing that these small 
systems are typically dominated by $>13$~Gyr old stars, which therefore 
formed at $z > 6$. In addition, many carbon-enhanced metal-poor (CEMP) 
stars have been found in ultra-faint dwarf galaxies \citep{norris10,
simon10,frebel15}, with respect to the more massive and more luminous 
``classical'' dwarf spheroidal galaxies \citep{skuladottir15}.
The high incidence of these peculiar stars, which have [C/Fe]$>$0.7
and were likely imprinted by the very first stars \citep{debennassuti14},
is an additional confirmation that ultra-faint galaxies are likely 
the living relics of star-forming mini-haloes from the pre-reionization epoch \citep{salvadori15}.

Accounting for the star-formation in mini-haloes is thus an essential
step to study the early phases of galaxy evolution and the impact 
on current stellar sample of ``second-generation" stars, which formed 
out of gas polluted by the first stars. In a previous paper, we studied 
the implication of the properties of the most iron-poor stars observed 
in the Galactic halo for the initial mass function (IMF) of the first Population~III (Pop~III) stars 
\citep{debennassuti14}. To this aim, we use the cosmological merger-tree 
code GAlaxy MErger Tree \& Evolution (\textsc{gamete}, \citealt{salvadori07,
salvadori08}) which was implemented to self-consistently account for the 
production and destruction of dust, and for a two-phase inter-stellar-medium 
\citep{valiante11,debennassuti14}. However, we did not account for star-formation 
in mini-haloes, as instead it was done in many other applications and 
further implementations of the model \citep[e.g.][]{salvadori09,salvadori10,
salvadori12,salvadori14,ferrara14,salvadori15,valiante16}. 

In \cite{debennassuti14}, we showed that dust-cooling is required to explain 
the existence of the most pristine, carbon-normal star at $Z\approx 10^{-4.5}\Zsun$
\citep{caffau11}. This implies that the transition from massive
Pop~III stars, to normal, Population~II (Pop~II) stars, is (also) driven
by thermal emission of collisionally excited dust and thus can occur at a critical 
dust-to-gas mass ratio ${\cal D}_{\rm cr}>4.4 \times 10^{-9}$ \citep{schneider02, schneider06, schneider12a}. 
Furthermore, we showed that CEMP-no stars (which show no r-/s- process elements, 
see Sec.~\ref{sec:obs}) are likely imprinted by the so-called ``primordial 
faint supernovae'' (SNe), i.e. Pop~III stars with typical progenitor masses of 
$\rm m_{PopIII}=[10-40]\Msun$ that during their evolution experience mixing and 
fallback and eject small amounts of $\rm Ni^{56}$, producing a faint lightcurve \citep{bonifacio03}. 
Our results pointed out that these pristine and relatively massive stars 
should dominate the early metal enrichment to successfully reproduce the 
observed fraction of Carbon-enhanced vs Carbon-normal stars. In particular, 
we put constraints on the Pop~III IMF, which should be limited to the mass 
range $\rm m_{PopIII}=[10-140] \Msun$. 
Note that state-of-the-art numerical simulations of the first cosmic 
sources performed by different groups predict disparate Pop~III star masses 
\citep[e.g.][for a recent review]{greif15},
with plausible mass ranges that can 
vary from sub-solar values \citep[e.g.][]{stacy16} up to $\rm m_{PopIII}
\approx 1000 \Msun$ \citep[e.g.][]{susa14,hirano14,hirano15,hosokawa16}.

In this paper, we investigate the impact of the Pop~III IMF on the 
properties of CEMP and C-normal stars by accounting for star-forming 
mini-haloes. To this aim, we further develop the model to catch the 
essential physics required to self-consistently trace the formation of 
stars in mini-haloes by including:
\begin{enumerate}
\item a star-formation efficiency that depends on the gas temperature, 
gas metallicity, and formation redshift of mini-haloes, and on the average 
value of the LW background;
\item a random sampling treatment of the IMF of Pop~III stars, which 
inefficiently form in mini-haloes;
\item a suppression of gas infall in mini-haloes born in ionized regions, 
using a self-consistent calculation of reionization.
\end{enumerate}
The paper is organized as follows: in Sec.~\ref{sec:model} we summarize the
main features of our cosmological model, mainly focusing on the new physics 
implemented here. In Sec.~\ref{sec:obs}, we briefly recap available observations
of very metal-poor stars in the Galactic halo, including the new findings.
Model results are presented in Sec.~\ref{sec:res}, where we show the effects
of different physical processes on the Galactic halo Metallicity Distribution Function (MDF)
and on the fraction 
of Carbon-enhanced vs Carbon-normal stars. Finally, in Sec.~\ref{sec:disc},  
we critically discuss the results and provide our conclusions.

%%%%%%%%%%%%%%%%%%%%%%%%%%%%%%%%%%%%%%%%%%%%%%%%%%%%%%%%%
\section{Description of the model}
\label{sec:model}
%%%%%%%%%%%%%%%%%%%%%%%%%%%%%%%%%%%%%%%%%%%%%%%%%%%%%%%%%
In this section we briefly summarize the main features of the semi-analytical
code \textsc{gamete} and describe in full detail the model implementations that
we made for the purpose of this work.

\textsc{gamete} is a cosmological merger tree model in the $\rm \Lambda$CDM 
framework\footnote{We assume a Planck cosmology with: $\rm h_0$=0.67, $\rm 
\Omega_b h^2$=0.022, $\rm \Omega_m$=0.32, $\rm \Omega_\Lambda$=0.68, $\rm 
\sigma_8$=0.83, $\rm n_s$=0.96 \citep{planckXVI}.} that traces the star-formation 
history and chemical evolution of Milky Way (MW)-like galaxies from redshift $z=20$ 
down to the present-day. The code reconstructs a statistical significant sample
of independent merger histories of the MW dark matter halo by using a binary 
Monte Carlo code, which is based on the Extended Press-Schechter (EPS) theory 
\citep[e.g.][]{bond91} and accounts for both halo mergers and mass accretion 
\citep{salvadori07}.
As shown by \cite{parkinson08}, merger trees obtained through the EPS 
formalism are in good agreement with the halo merger histories of N-body 
simulations \citep[e.g. the Millennium simulation,][]{springel05} and the same 
is true for the specific case of the MW Galaxy \citep[e.g.][Fig.~2.3]{salvadori09Phd}.
Since semi-analytical models can rapidly run on merger trees, this approach 
enables us to efficiently explore the effects of different model parameters on 
a large sample of possible assembly histories of our Galaxy, which is unknown. 
Hence, it is the perfect statistical tool to study the impact of the Pop~III 
IMF on the observed properties of present-day stars while accounting for the
uncertainties induced by different assembling histories. To reconstruct the
possible MW merger histories, we assume that the Galaxy is embedded in a DM 
halo of mass $M_{\rm MW}=10^{12}\, \Msun$ at $z=0$, which is in agreement with 
current measurements and uncertainties: $1.26^{+0. 24}_{-0.24}\times 10^{12} \,\Msun$ \citep{mcmillan11}, $0.9^{+0.4}_{-0.3} \times 10^{12} 
\, \Msun$ \citep{kafle12}, $0.80^{+0.31}_{-0.16}\times 10^{12} \, \Msun$ (\citealt{kafle14}, 
see also \citealt{wang15} for the different methods adopted to infer the DM mass).
The 50 possible MW merger histories we reconstruct resolve mini-haloes down to a virial 
temperature $T_{\rm vir} = 2\times 10^3 K$ \citep[e.g. see][]{debennassuti14}.
We use 50 possible MW merger histories because we find that the results 
converge with this number of realizations, thus a larger number would not provide
any improvement in our findings.

The star-formation and chemical evolution history of the MW is then traced 
along the merger trees. We assume that the gas inside dark matter haloes can 
be converted into stars at a rate $\psi(t)=\epsilon_* M_{\rm ISM}(t)/t_{\rm dyn}(t)$, 
where $\epsilon_*$ is the star-formation efficiency, $t_{\rm dyn}(t)$ the dynamical
time-scale, and $M_{\rm ISM}(t)$ the total mass of gas into the 
interstellar medium (ISM), which is regulated by star formation, SN-driven outflows,
and by a numerically calibrated infall rate \citep{salvadori08}. 
Due to less efficient gas cooling, the star formation efficiency in mini-haloes ($\rm T_{\rm vir} < 10^4 K$),
$\epsilon_{\rm MH}$, is smaller than that of Ly$\alpha$-cooling haloes ($\rm T_{\rm vir} \ge 10^4 K$),
$\epsilon_*$ (see Sec.~\ref{sec:cool}). 

Once stars are formed, we follow their subsequent 
evolution using mass- and metallicity-dependent lifetimes \citep{raiteri1996}, 
metal \citep{hoek1997,woosley95,heger02} 
and dust \citep{schneider04,bianchi07,marassi14,marassi15} yields. 
Chemical evolution is
followed in all star-forming haloes and in the surrounding MW 
environment enriched by SN-driven outflows
regulated by a wind efficiency $\epsilon_w$, 
which represents the fraction of SN explosion energy
converted into kinetic form \citep{salvadori08}. While $\epsilon_w$ is assumed
to be the same for all dark matter haloes, the ejected mass is evaluated by comparing 
the SN kinetic energy with the halo binding energy.
Following Eq.~9 of \cite{salvadori08}, we thus have:
\begin{equation}
\frac{dM_{ej}}{dt} \propto \frac{\epsilon_w}{v_{circ}^2}
\end{equation}
where $\rm v_{circ}$ is the halo circular velocity.

As discussed in \cite{debennassuti14}, the evolution of the ISM in each 
progenitor halo is described as follows:

\begin{itemize}
\item we assume the ISM to be characterized by two phases: ({\it i}) a cold, {\it dense} 
phase that mimic the properties of molecular clouds (MCs). In this dense phase 
star-formation occurs and dust grains accrete gas-phase metals. ({\it ii}) A hot, {\it diffuse} 
phase that exchanges mass with the MW environment through gas infall and SN winds. 
In this diffuse phase, SN reverse shocks can partially destroy newly produced dust 
grains \citep{bianchi07, marassi15, bocchio16}.

\item Chemical evolution is described separately in the two phases, 
but the mass exchange between the dense and diffuse ISM is taken into account,
through the condensation of the diffuse phase and the 
dispersion of MCs which return material to the diffuse phase. These processes 
are regulated by two additional free parameters, as described in Section 
2.1 of \cite{debennassuti14}.

\item The transition from massive, Pop~III stars to 
``normal'' Pop~II stars is assumed to be driven by dust 
grains \citep[e.g.][]{schneider02} and to occur when the dust-to-gas 
mass ratio in the dense phase exceeds the critical value 
$\mathcal{D}_{cr} = 4.4\times 10^{-9}$ \citep{schneider12b}

\item Pop~III stars form with masses in the range $ [10-300]\, \Msun$ according to a Larson-type IMF:
\begin{equation}
\Phi(m) = \frac{dN}{dm} \propto m^{\alpha-1} \, {\rm exp}\Biggl(-\frac{m_{\rm ch}}{m}\Biggr)
\label{eq:larson}
\end{equation}
with $m_{\rm ch} = 20 \, \Msun$ and $\rm \alpha = 1.35$  \citep{debennassuti14}. We assume that 
chemical enrichment (metals and dust) is driven by Pop~III stars with masses 
$[10-40] \, \Msun$, that explode as faint SNe, 
and by stars with masses $[140-260] \, \Msun$, that explode as 
Pair Instability SNe \citep[PISNe, e.g.][]{heger02,schneider04,takahashi16}.
We recall that these ``faint SNe'' experience mixing and fallback during 
their evolution \citep{bonifacio03}. As a consequence, when they explode as SNe, they eject small amounts of $\rm Ni^{56}$ 
(and hence Fe) with respect to other light elements, such as C 
\citep[e.g.][]{umeda03,iwamoto05,tominaga07,marassi14,ishigaki14}. The Fe and C yields adopted 
in this work \citep{marassi14,marassi15} are in agreement with the findings 
of other groups that assumed similar faint SN models \citep[e.g.][] {iwamoto05,tominaga07,ishigaki14}.
Note that other scenarios have been proposed for the evolution of Pop~III stars 
in this relatively low mass range, such as the rapidly rotating ``spinstars'' 
\citep[e.g.][]{meynet06,maeder15}. Still, even in these cases, the yields of Fe 
and C are very similar to those produced by the faint SNe adopted in this work.

\item Pop II stars form according to a Larson IMF with 
$m_{\rm ch} = 0.35 \, \Msun$, $\rm \alpha = 1.35$, and masses in the range 
$[0.1-100]\, \Msun$. Their contribution to chemical enrichment
is driven by Asymptotic Giant Branch (AGB) stars 
($2 \, M_\odot \leq m_{\rm PopII} \leq 8 M_\odot$) and by
ordinary core-collapse SNe ($8 \, M_\odot < m_{\rm PopII} \leq 40 \, M_\odot$).
\end{itemize}

We refer the reader to the aforementioned papers for a detailed description of 
the basic features of the model. In the next subsections we illustrate how we 
implemented star-formation in $\rm H_2$ cooling mini-haloes.

%%%%%%%%%%%%%%%%%%%%%%%%%%%%%%%%%%%%%%%%%%%%%%%%%%%%%%%%%
%%% Gas cooling
\begin{figure}
\includegraphics[width=1.6\columnwidth]{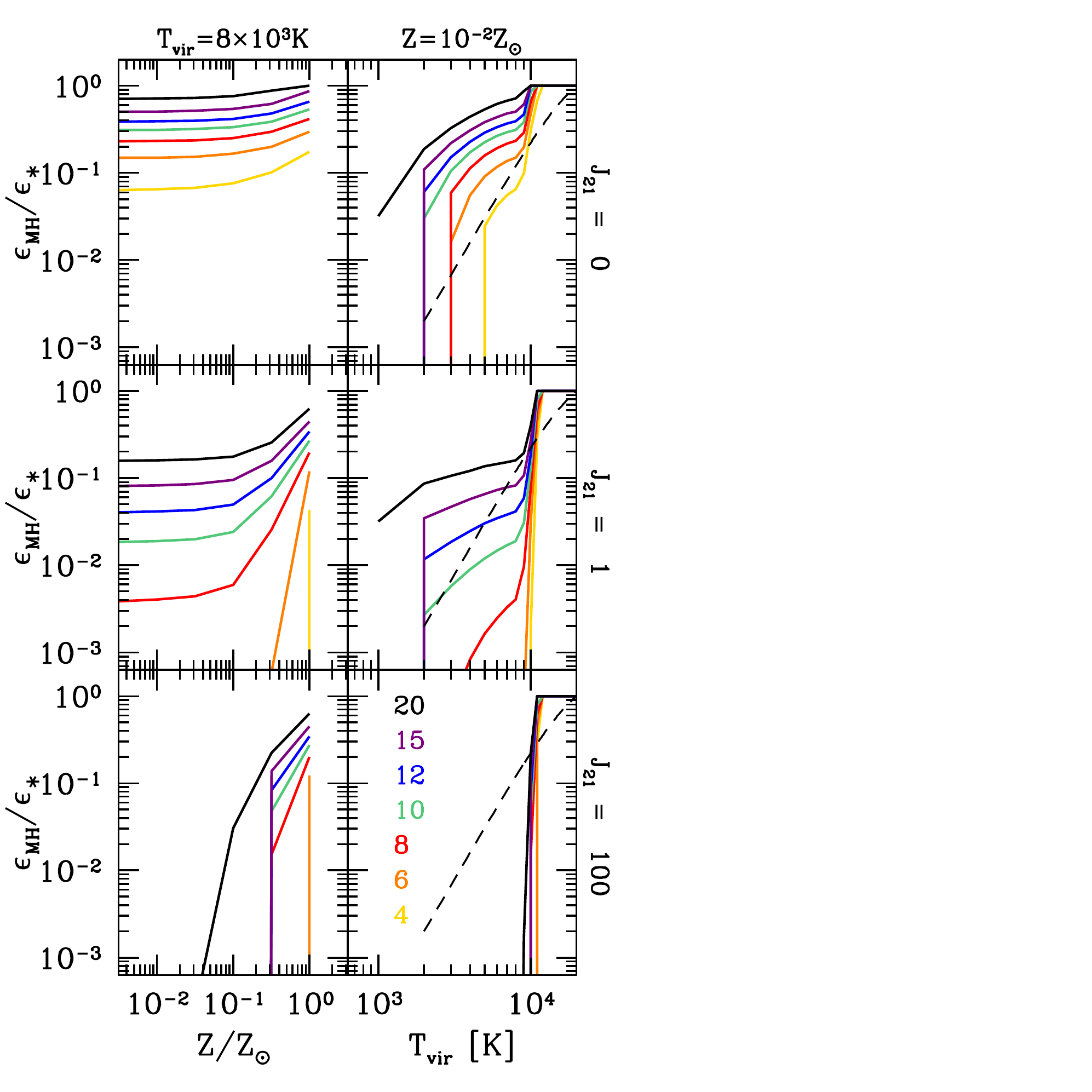}
\caption{The ratio between the star-formation efficiency in mini-haloes, $\epsilon_{\rm MH}$,
and the constant efficiency of Lyman-$\alpha$ cooling haloes $\epsilon_{*}$ (see text). The
colored lines represent different mini-haloes formation redshifts ($z = 20, 15, 12, 10, 8, 6, 4$, from top to bottom,
as explained in the legenda). Upper, middle and lower panels show the results assuming 
$J_{21} = J_{LW}/(10^{-21} \rm erg/cm^2/s/Hz/sr)$ = 0, 1 and 100,
respectively. {\it Left panels}: dependence on the gas metallicity
for a fixed $T_{\rm vir} = 8 \times 10^3$~K; at Z $< 10^{-2.5} Z_\odot$ 
the ratio $\epsilon_{MH}/\epsilon_\ast$ is constant. 
{\it Right panels}: dependence on the virial temperature, 
for a fixed $\rm Z = 10^{-2} Z_\odot$. 
The dashed lines show the $z$-independent relation used in 
\citet{salvadori09,salvadori12}.
} 
\label{fig:cooling}
\end{figure}
%%%%%%%%%%%%%%%%%%%%%%%%%%%%%%%%%%%%%%%%%%%%%%%%%%%%%%%%%

%%%%%%%%%%%%%%%%%%%%%%%%%%%%%%%%%%%%%%%%%%%%%%%%%%%%%%%%%
\subsection{Gas cooling in mini-haloes}
\label{sec:cool}

The gas cooling process in mini-haloes relies on the presence of molecular 
hydrogen, H$_2$, which can be easily photo-dissociated by photons in the LW band. 
Several authors have shown that ineffective cooling by $\rm H_2$
molecules limits the amount of gas that can be converted into stars, with 
efficiencies that decrease proportional to $T^3_{\rm vir}$ \citep[e.g.][]{madau01,
okamoto08}. For this reason, models that account for mini-haloes, typically 
assume that in these small systems the star-formation efficiency is reduced 
with respect to more massive Lyman-$\alpha$ cooling haloes, and that the ratio
between the two efficiencies is a function of the virial temperature,
$\epsilon_{\rm MH}/\epsilon_*=2[1+(T_{\rm vir}/2\times 10^4\,{\rm K})^{-3}]^{-1}$ 
\citep[e.g.][]{salvadori09,salvadori12}.
This simple relation is illustrated in the right panels of Fig.~\ref{fig:cooling} (dashed lines).

However, the ability of mini-haloes to cool down their gas not only depends
on $T_{\rm vir}$, but also on their formation redshift, on the gas metallicity, 
and on the LW flux, $J_{\rm LW}$, to which these systems are exposed. In this
work, we compute the mass fraction of gas that is able to cool in one dynamical 
time following a simplified version of the chemical evolution model of \citet{omukai12}.
For a more detailed description of the model and of the gas-cooling processes that
we have considered, we refer the reader to the Appendix A of \cite{valiante16}\footnote{
Here we correct the derivation of the cooling time (Eq.~A2 in \citealt{valiante16}) 
by dividing Eq.~19 of \cite{madau01}
for $\Omega_m$. This enhances the star formation efficiencies 
shown in their Fig.~A1-A4 (see Fig.~\ref{fig:cooling}).}.

In Fig.~\ref{fig:cooling} we show the resulting efficiencies, $\epsilon_{\rm MH}$, 
normalized to $\epsilon_*$ 
for different gas metallicities (left), mini-halo virial temperatures 
(right) and LW flux. 
The different lines show different mini-halo formation redshifts.
When the LW background is neglected, i.e. 
$J_{21} = J_{LW}/(10^{-21} \rm erg/cm^2/s/Hz/sr) = 0$ (upper panels), 
we can clearly see the
dependence of $\epsilon_{\rm MH}/\epsilon_*$ on:
({\it i}) the formation redshift, ({\it ii}) the gas metallicity (left), and ({\it iii}) the virial temperature (right). 
As a general trend, we find that the smaller are these quantities, the lower is the gas-cooling efficiency of mini-haloes. 
At $T_{\rm vir} \ge 10^4$~K, Lyman-$\alpha$ cooling becomes efficient and 
$\epsilon_{\rm MH}/\epsilon_* = 1$. When $J_{21} \sim 1$ (middle panels), 
gas cooling is partially suppressed due to H$_2$ photo-dissociation.
When $z > 8$, the gas is dense enough to self-shield against the external 
LW background, and $\epsilon_{\rm MH}$ decreases only by 1 dex with respect 
to the $J_{\rm 21} = 0$ case. When $z \le 8$, gas cooling is completely 
suppressed in $T_{\rm vir} < 10^4$~K systems, unless they are already 
enriched to $Z \sim Z_{\odot}$. This is consistent with the minimum halo 
mass to form stars assumed in \cite{salvadori09,salvadori12}. 
Similarly, when mini-haloes are exposed to a stronger LW flux, $J_{21} 
\sim 100$ (lower panels), gas cooling is allowed only at $z>8$ in highly 
enriched mini-haloes, which have $Z \gtrsim 10^{-1.5}Z_{\odot}$. In what 
follows, we will use the results in Fig.~\ref{fig:cooling} to self-consistently 
compute the star-formation efficiency of mini-haloes in our merger trees.

%%% Stochastic IMFs %%%%%%%%%%%%%%%%%%%%%%%%%%%%%%%%%%%%%%%%%%%%
\begin{figure*}
\centering
\includegraphics[trim=0 1.cm 0 13cm, clip=true, width=18cm]{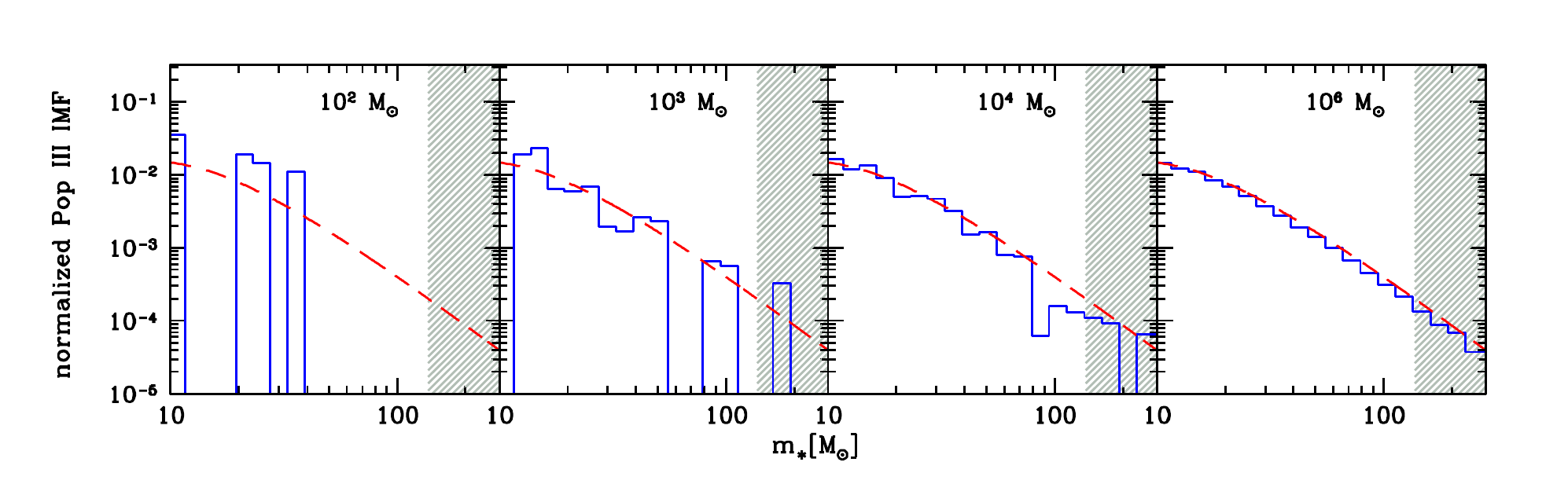}
\caption{Comparison between the intrinsic IMF of Pop~III stars (red dashed lines) and 
the effective mass distribution resulting from the random sampling procedure (blue histograms).
In each panel, the IMF is normalized to the total stellar mass formed, with 
$M^{\rm tot}_{\rm PopIII} = (10^2, 10^3, 10^4, 10^6)\, \Msun$ (from left to right). 
The grey shaded areas indicate the progenitor mass range of PISNe (see text).}
\label{fig:imf}
\end{figure*}
%%%%%%%%%%%%%%%%%%%%%%%%%%%%%%%%%%%%%%%%%%%%%%%%%%%%%%%%%
\subsection{Stochastic Pop III IMF}
\label{sec:stocimf}
%%%%%%%%%%%%%%%%%%%%%%%%%%%%%%%%%%%%%%%%%%%%%%%%%%%%%%%%%

The mass of newly formed stars in a given dark matter halo depends 
on the gas mass and star formation efficiency. In low-mass systems 
the available mass of gas can be strongly reduced with respect to the
initial mean cosmic value, $\Omega_{\rm b}/\Omega_{\rm M}$, because of 
SN-driven outflows and radiative feedback processes \citep[e.g.][]{salvadori09}. 
Furthermore, due to the reduced cooling efficiency mini-haloes (see Fig.~\ref{fig:cooling}),
the total stellar mass formed in each burst is small, $M^{\rm tot}_{\rm PopIII} < 10^4 \, M_\odot$.
In these conditions, the resulting stellar mass spectrum will be affected by the incomplete
sampling of the underlying stellar IMF.
This effect can be particularly 
relevant for Pop~III stars, which have masses in the range $[10-300] \Msun$ 
\citep[e.g.][]{hirano14}. 
Hence, we adopt 
a random-selection procedure, the result of which is illustrated in
Fig.~\ref{fig:imf}.
 
The stars produced during each burst of star-formation are randomly selected 
within the IMF mass range, and they are assumed to form with a probability that 
is given by the IMF normalized to the total mass of stars formed in each burst.
In Fig.~\ref{fig:imf} we show the comparison between the randomly-selected and the 
intrinsic IMF for  
total stellar masses of $M^{\rm tot}_{\rm PopIII} = (10^2, 10^3, 10^4, 10^6)\, \Msun$. 
When $M^{\rm tot}_{\rm PopIII} \gtrsim 10^5 \, \Msun$ the overall mass 
range can be fully sampled and the intrinsic IMF is well reproduced. On the other
hand, the lower is the stellar mass formed, the worse is the match between the 
sampled and the intrinsic IMF. In particular, when $M^{\rm tot}_{\rm PopIII} \leq 
10^3 \, \Msun$, it becomes hard to form  Pop~III stars with 
$\rm m_{PopIII}=[140-260] \Msun$
(PISN progenitor mass range). 
We find that while almost 100\% of the haloes forming $\rm 10^4 \Msun$ Pop~III stars host 
$\approx$ 9 PISNe each, only 60\% of the haloes producing $\rm 10^3 \Msun$ of 
Pop~III stars host {\it at most} 1 PISN. Thus, we find that the number
of PISN is naturally limited by the incomplete sampling of the stellar IMF. 

%%%%%%%%%%%%%%%%%%%%%%%%%%%%%%%%%%%%%%%%%%%%%%%%%%%%%%%%%
\subsection{Radiative feedback}
\label{sec:rad}
%%%%%%%%%%%%%%%%%%%%%%%%%%%%%%%%%%%%%%%%%%%%%%%%%%%%%%%%%
To self-consistently account for the effect of radiative feedback processes 
acting on mini-haloes, we compute the amount of radiation in the Lyman Werner 
(LW, 11.2-13.6 eV) and ionizing ($>$ 13.6 eV) bands produced by star-forming 
haloes along the merger trees. 
We take time- and metallicity-dependent UV luminosities from \cite{bruzual03} for Pop~II 
stars and from \cite{schaerer02} for Pop~III stars\footnote{We use the photon
luminosities from Table~4 of \cite{schaerer02} for metal-free stars.}. 

\begin{figure}
\centering
\includegraphics [trim=1.5cm 2cm 0cm 9cm, clip=true,width=9.cm]{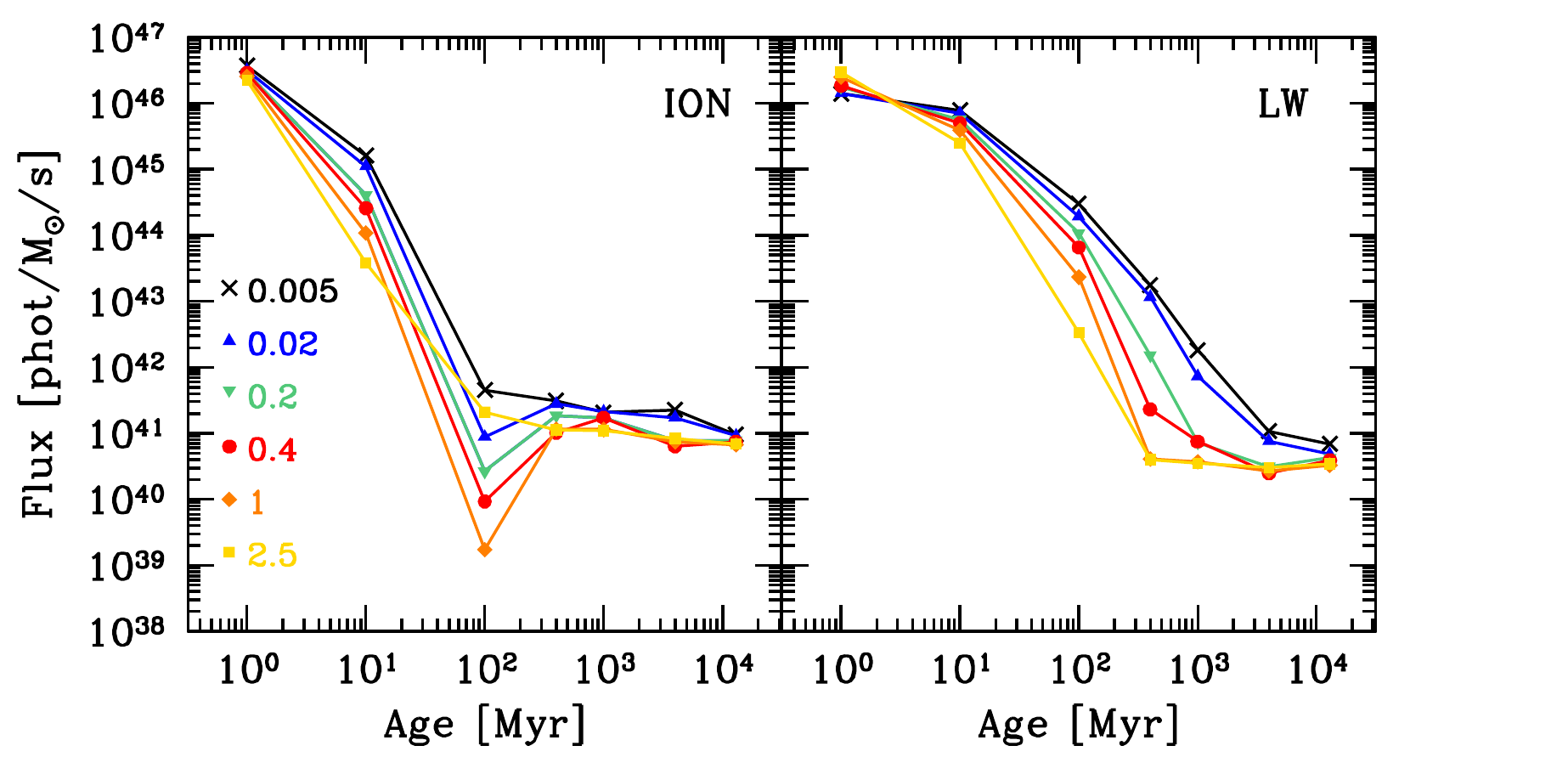}
\caption{Ionizing (left) and LW (right) photon luminosities per unit stellar mass 
of Pop~II stars as a function of stellar age \citep{bruzual03}. Different colors 
refer to different metallicities as shown by the labels: $\rm Z/\Zsun$ = 0.005, 0.02, 
0.2, 0.4, 1, 2.5 (black crosses, blue up-triangles, green down-triangles, red dots, orange diamonds, gold squares).}
\label{fig:flux}
\end{figure}

Input values used for Pop~II stars of different stellar metallicities are shown in Fig.~\ref{fig:flux}
as a function of stellar ages.
As expected, the UV luminosity is largely dominated
by young stars and depends on stellar metallicity, as more
metal poor stars have harder emission spectra. 
In the first 10 Myr, the ionizing photon rate drops  from 
$> 2\times 10^{46} \, {\rm phot/s/}M_\odot$ to $< 2\times 10^{45} 
\, {\rm phot/s}/M_\odot$, while the rate of LW photons 
remains approximately constant at $\sim 10^{46} \, {\rm phot/s/}M_\odot$. 
Therefore, although initially lower, after 10 Myr the rate of
LW photons emitted becomes higher than that of ionizing photons.

In the following, we describe how we compute the LW and ionizing 
background and how we account for reionization of the MW environment. 
Given the lack of spatial information, we assume that the time-dependent LW and
ionizing radiation produced by stars build-up a homogeneous background, 
which - at each given redshift - affects all the (mini-)haloes in the same way.\\

%%%%%%%%%%%%%%%%%%%%%%%%%%%%%%%%%%%%%%%%%%%%%%%%%%%%%%%%%
\subsubsection{LW background}
\label{sec:LW}

The cumulative flux observed at a given frequency $\nu_{\rm obs}$ and redshift $z_{\rm obs}$ can 
be computed by accounting for all stellar populations that are still evolving at 
$z_{\rm obs}$ \citep{haardt1996}:
\begin{equation}
J(\nu_{\rm obs}, z_{\rm obs}) = (1+z_{\rm obs})^{3} \frac{c}{4\pi} \int_{z_{\rm obs}}^{\infty} dz \, \, \bigg|\frac{dt}{dz}\bigg| \, \epsilon(\nu',z) \, e^{{-\tau(\nu_{\rm obs}, z_{\rm obs}, z)}}
\label{eq:flux}
\end{equation}
\noindent
where 
\begin{equation}
\bigg|\frac{dt}{dz}\bigg| = \{ H_0 (1+z) [\Omega_{\rm m} (1+z)^3+\Omega_\Lambda]^{1/2} \}^{-1}
\label{eq:dtdz},
\end{equation}
\noindent
$\nu'=\nu_{\rm obs}(1+z)/(1+z_{\rm obs})$, $\epsilon(\nu',z)$ is the comoving emissivity at 
frequency $\nu'$ and redshift $z$, and $\tau(\nu_{\rm obs}, z_{\rm obs}, z)$ is the MW environment optical 
depth affecting photons emitted at redshift $z$ and seen at redshift $z_{\rm obs}$ at frequency 
$\nu_{\rm obs}$ \citep{ricotti01}.

To obtain the average flux in the LW band, we  integrate Eq.~(\ref{eq:flux}) between
$\nu_{\rm min} = 2.5\times10^{15}\, {\rm Hz}$ (11.2~{\rm eV}) and $\nu_{\rm max} = 3.3\times10^{15}~{\rm Hz}$ (13.6~eV):
\begin{equation}
J_{\rm LW}(z_{\rm obs}) = \frac{1}{\nu_{\rm max}-\nu_{\rm min}}\int_{\nu_{\rm min}}^{\nu_{\rm max}} d\nu_{\rm obs} \,\, J(\nu_{\rm obs}, z_{\rm obs}).
\end{equation}
\noindent
Following \cite{ahn09}, we write the mean attenuation in the LW band as:
\begin{equation}
e^{-\tau(z_{\rm obs}, z)}= \frac{\int_{\nu_{\rm min}}^{\nu_{\rm max}} e^{-\tau(\nu_{\rm obs}, z_{\rm obs}, z)}\,d\nu_{\rm obs}}{\int_{\nu_{\rm min}}^{\nu_{\rm max}}d\nu_{\rm obs}},
\end{equation}
\noindent
so that,
\begin{equation}
J_{\rm LW}(z_{\rm obs}) = (1+z_{\rm obs})^{3}\,\, \frac{c}{4\pi} \int_{z_{\rm obs}}^{z_{\rm screen}} \, dz \, \bigg|\frac{dt}{dz}\bigg| \, \epsilon_{\rm LW}(z) \, e^{-\tau(z_{\rm obs}, z)}.
\label{eq:fluxLW}
\end{equation}
\noindent
where $\epsilon_{\rm LW}(z)$ is the emissivity in the LW band.
Eq.~\ref{eq:fluxLW} is integrated up to $z_{\rm screen}$, 
that is the redshift above which photons emitted in the LW band are redshifted out of the band (the so-called ``dark screen'' effect).

In the top panel of Fig.~\ref{fig:radiation}, we show the redshift 
evolution of the LW background predicted by the fiducial model (see 
Sec.~\ref{sec:prop} and \ref{sec:res}). We find that at $z \sim 18$ 
the average LW background is already $J_{21} \approx 10$, thus 
strongly reducing the star-formation in $T_{\rm vir} \, < \, 10^4$~K 
mini-haloes. The point shown at $z~=~0$ is the average LW background 
in the Galactic ISM, and it has been computed from Eq.~(2) 
of \cite{sternberg14}. The value
refers to a LW central wavelength of 1000 $\AA$, while the errorbar 
represents the variation interval of $J_{21}$ corresponding to 
all wavelegths in the LW band (i.e. from 912 $\AA$ to 1108 $\AA$). 
Although the $J_{21}$ background we have obtained is consistent with 
this data, at high-$z$ our calculations likely overestimate $J_{21}$. 
In fact, at any given redshift all LW photons are assumed to escape 
from star-forming galaxies but they are retained into the MW environment. 
Yet, the mean free path of LW photons 
\citep[$\approx 10$~Mpc, e.g.][]{haiman97} is larger than the physical 
radius of the Milky Way.
In particular, we assume that the MW and its progenitor haloes
evolve in a comoving volume $\approx 5$~Mpc$^3$,
which we estimate at the turn-around radius \cite[see][]{salvadori14}.
Hence our code simulates a biased region of the Universe, the high density
fluctuation giving rise to the MW and its progenitors. For this reason, the 
LW flux we obtain is consistent with that found by \cite{crosby13} who consider 
a similar volume. On the other hand, the predicted $J_{21}$ is larger than 
those obtained by cosmological simulations of the Universe, which account 
for larger volumes \citep[e.g.][]{ahn09,oshea15,xu16}. However, as we will 
show in Sec.~\ref{sec:res}, such a strong difference in the LW background 
have a very small impact on our results.

%%%%%%%%%%%%%%%%%%%%%%%%%%%%%%%%%%%%%%%%%%%%%%%%%%%%%%%%%%%%%%%%%%%%%%%%%%%%%%%%%%%
%%% Radiative feedback
%%%%%%%%%%%%%%%%%%%%%%%%%%%%%%%%%%%%%%%%%%%%%%%%%%%%%%%%%%%%%%%%%%%%%%%%%%%%%%%%%%%%
\subsubsection{Reionization}
\label{sec:reion}
%%%%%%%%%%%%%%%%%%%%%%%%%%%%%%%%%%%%%%%%%%%%%%%%%%%%%%%%%%%%%%%%%%%%%%%%%%%%%%%%%%%%
Photons with energies $>$ 13.6 eV are responsible for hydrogen (re)ionization. 
At any given redshift, we compute the ionizing photon rate density, 
$\dot{n}_{\rm ion}(z)$, by summing the ionizing luminosities over all active 
stellar populations and dividing by the MW volume, $\approx 5$~Mpc$^3$.
The filling factor $\rm Q_{\rm HII}$, i.e. the fraction of MW volume which has been reionized 
at a given observed redshift $z_{\rm obs}$, is computed following \cite{barkana01}:
\begin{equation}
Q_{\rm HII}(z_{\rm obs}) = \frac{f_{\rm esc}}{n_{\rm H}^0}\int_{z_{\rm obs}}^{z_{\rm em}} \, dz \, \bigg| \frac{dt}{dz} \, \bigg| \, \dot{n}_{\rm ion} \, e^{F(z_{\rm obs},z)}
\label{eq:reion}
\end{equation}
\noindent
with
\begin{equation}
F(z_{\rm obs},z)=-\alpha_{\rm B} \, n_{\rm H}^0 \int_{z_{\rm obs}}^{z} dz' \, \bigg| \frac{dt}{dz'} \bigg| \, C(z') \, (1+z')^3
\label{eq:reionfactor}
\end{equation}
\noindent
where $f_{\rm esc}$ is the escape fraction of ionizing photons, $F(z_{\rm obs},z)$ 
accounts for recombinations of ionized hydrogen, and $n_{\rm H}^0$ is the present-day
hydrogen number density in the MW environment,  which is $\approx 5$ times larger 
than in the IGM. In Eq.~\ref{eq:reionfactor}, 
$\alpha_B$ is the case B recombination coefficient\footnote{We assume $\rm \alpha_B~=~2.6 
\times 10^{-13}~cm^3/s$, which is valid for hydrogen at $ T~=~10^4$~K, e.g.
Maselli et al. (2003).}, and $C(z)$ is the redshift-dependent clumping factor, which is 
assumed to be equal to $C(z)~=~17.6\,e^{(-0.10\,z+0.0011\,z^2)}$ 
\citep{iliev05}. 

\begin{figure}
\centering
\includegraphics [trim=6.cm 1.cm 5.5cm 1.cm, clip=true, width=0.8\columnwidth]{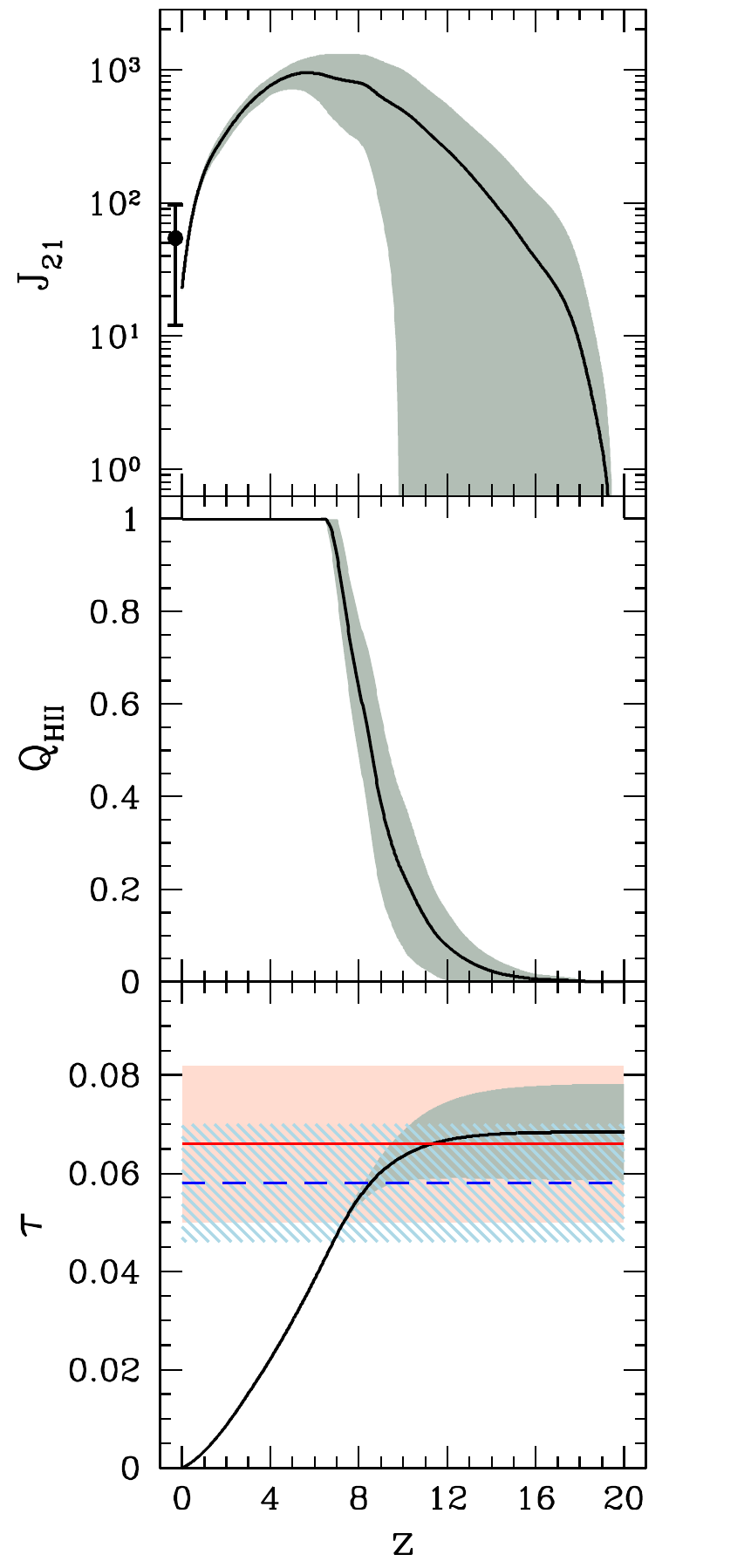}
\caption{Redshift evolution of the LW flux ({\it upper panel}), the volume filling factor of ionized 
regions ({\it middle panel}) and the corresponding Thomson scattering optical depth ({\it lower panel})
predicted by the fiducial model and 
obtained by averaging over 50 MW merger histories (solid lines). The shaded areas show the 
1-$\sigma$ dispersion among different realizations. The point with error bar at $z=0$ 
represents an estimate of the specific intensity of the isotropic UV radiation field in the 
Galactic ISM \citep{sternberg14}.
The red-solid horizontal line shows the Planck measurement \citep{planck2015} 
with 1-$\sigma$ dispersion (horizontal filled shaded area). 
The more recent measurement \citep{planck2016} is plotted as a blue-dashed horizontal line 
with 1-$\sigma$ dispersion (horizontal striped-dashed shaded area).
}
\label{fig:radiation}
\end{figure}

Following \citet{salvadori14}, we assume an escape fraction for ionizing photons 
$f_{\rm esc}$=0.1, which provides a reionization history that is complete by $z 
\sim 6.5$, fully consistent with recent data. This is illustrated in the 
lower panels of Fig.~\ref{fig:radiation}, where we show the redshift evolution of 
the volume filling factor of ionized regions and the corresponding Thomson 
scattering optical depth. In addition, \citet{salvadori14} show that the same 
reionization history appears consistent with the star-formation histories of 
dwarf satellites of the MW.

%%%%%%%%%%%%%%%%%%%%%%%%%%%%%%%%%%%%%%%%%%%%%%%%%%%%%%%%%%%%%%%%%%%%%%%%
%%% Radiative feedback
%%%%%%%%%%%%%%%%%%%%%%%%%%%%%%%%%%%%%%%%%%%%%%%%%%%%%%%%%%%%%%%%%%%%%%%

%%%%%%%%%%%%%%%%%%%%%%%%%%%%%%%%%%%%%%%%%%%%%%%%%%%%%%%%%%%%%%%%%%%%%%%%%%%%%%%%%%%%%%%%
\subsection{Model calibration}
\label{sec:prop}
%%%%%%%%%%%%%%%%%%%%%%%%%%%%%%%%%%%%%%%%%%%%%%%%%%%%%%%%%%%%%%%%%%%%%%%%%%%%%%%%%%%%
The free parameters of the model are assumed to be the same for all dark
matter haloes of the merger trees and they are calibrated by comparing 
the ``global properties'' of the MW as observed today with the average 
mean value over 50 independent merger histories. In particular, we match the 
observed mass of stars, metals, dust and gas, along with the star formation 
rate \citep[see Table~1 of][]{debennassuti14}. All these observed properties
are simultaneously reproduced by assuming the same model free parameters of 
\cite{debennassuti14}, Table~2. By looking at the predicted evolution of such
a global properties we can see that at $z = 0$ the 1-$\sigma$ dispersion among 
different merger histories becomes smaller than the observational errors 
(Fig.~4 in de Bennassuti et al. 2014). This implies that the global properties 
are well reproduced not only by the average mean values, but also by each single 
merger history.
%%%%%%%%%%%%%%%%%%%%%%%%%%%%%%%%%%%%%%%%%%%%%%%%%%%%%%%%%%%%%%%%%%%%%%%%
%%% MDF: comparisons and reference %%%
\begin{figure}
\includegraphics[width=1.0\columnwidth]{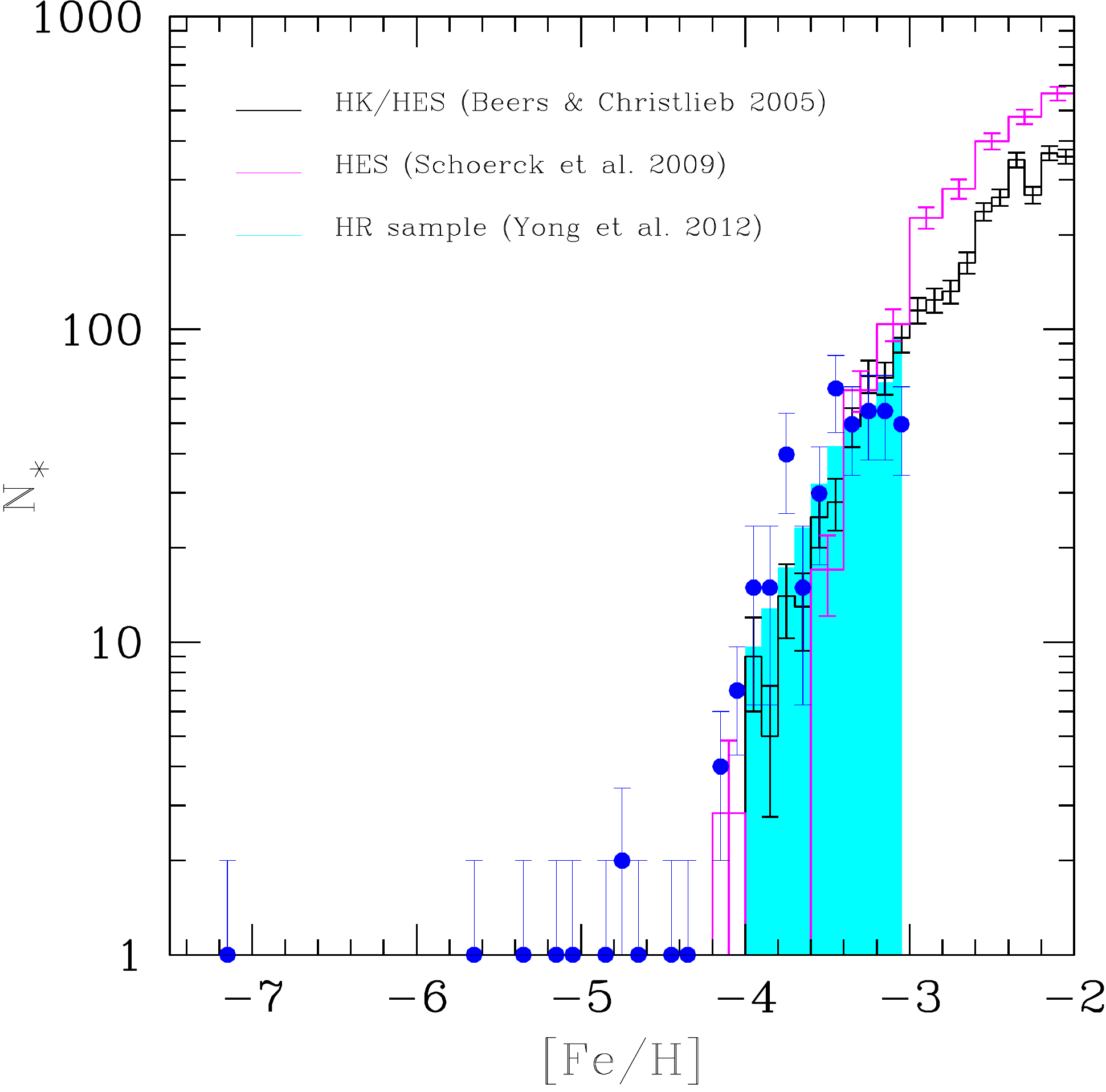}
\caption{The Galactic halo MDFs obtained using different data-sets 
and normalized to the same number of stars at [Fe/H]$\, \leq -3$. 
Histograms show the result from:
{\it black} - the medium resolution HK and HES surveys \citep{beers05}; 
{\it magenta} - the HES survey corrected for observational biases 
and incompleteness \citep{schoerck09}; {\it cyan shaded} -
the homogeneous sample of high-resolution (HR) spectroscopic data 
from \cite{yong13}, corrected to account for 
incompleteness and observational errors. The {\it blue points} show 
the uncorrected sample by \cite{yong13} to which we added HR data for stars with 
[Fe/H]$<-4$, taken from recent literature (see text for details 
and references). Errorbars represent poissonian errors.} 
\label{fig:MDF}
\end{figure}
%%%%%%%%%%%%%%%%%%%%%%%%%%%%%%%%%%%%%%%%%%%%%%%%%%%%

%%%%%%%%%%%%%%%%%%%%%%%%%%%%%%%%%%%%%%%%%%%%%%%%%%%%%%%%%%%%%%%%%%%%%%%%%%%%%%%%%%%%%%%%
\section{Observations: very metal-poor stars}
\label{sec:obs}
%%%%%%%%%%%%%%%%%%%%%%%%%%%%%%%%%%%%%%%%%%%%%%%%%%%%%%%%%%%%%%%%%%%%%%%%%%%%%%%%%%%%%%%%%
During the last decades, surveys looking for very metal-poor 
stars, with [Fe/H]$<-2$, have explored the stellar halo of our 
Galaxy. One of the main outcomes of these surveys has been 
the determination of the MDF, namely the number of stars as 
a function of their iron abundance, [Fe/H], which is used as a 
metallicity tracer. 
Early determinations of the MDF by \cite{ryan_norris91}
have shown that the MDF peaks around [Fe/H]$=-1.6$,
with wings from solar abundances down to [Fe/H]$\approx -3$,
although one star at [Fe/H]$\approx -4$ was already
discovered thanks to serendipitous detection \citep{bessell_norris84}.
Spectroscopic follow-up of the HK survey \citep{beers85} by \cite{molaro90},
\cite{bonifacio90} and \cite{primas94} showed that the tail extended down to 
[Fe/H]$\approx-4.0$, and led to the determination of the first 
Carbon-enhanced metal-poor stars \citep{norris97,bonifacio98}.
More recent observations by the HK and the Hamburg/ESO Survey 
(HES, e.g. Christlieb et al. 2008) have confirmed the presence of 
the MDF peak at [Fe/H]$\approx-1.6$ and they have lead to the 
identification of some hundreds of new stars at [Fe/H]$<-3$ (see 
Fig.~\ref{fig:MDF}). These ``extremely metal-poor stars'' are of key 
importance to understand the early chemical enrichment processes. 
In particular, it has been shown that the shape of the low-Fe tail of 
the Galactic halo MDF can shed new light on the properties of the 
first stellar generations, and on the physical processes driving the 
transition from massive Pop~III stars to normal Pop~II stars
 \citep{tumlinson06,salvadori07,komiya07,salvadori10,tumlinson10,
debennassuti14, hartwig15, komiya16}.  

%%%%%%%%%%%%%%%%%%%%%%%%%%%%%%%%%%%%%%%%%%%%%%%%%%%%%%%%
%%% FCEMP: comparisons and reference %%%
\begin{figure}
\includegraphics[width=1.0\columnwidth]{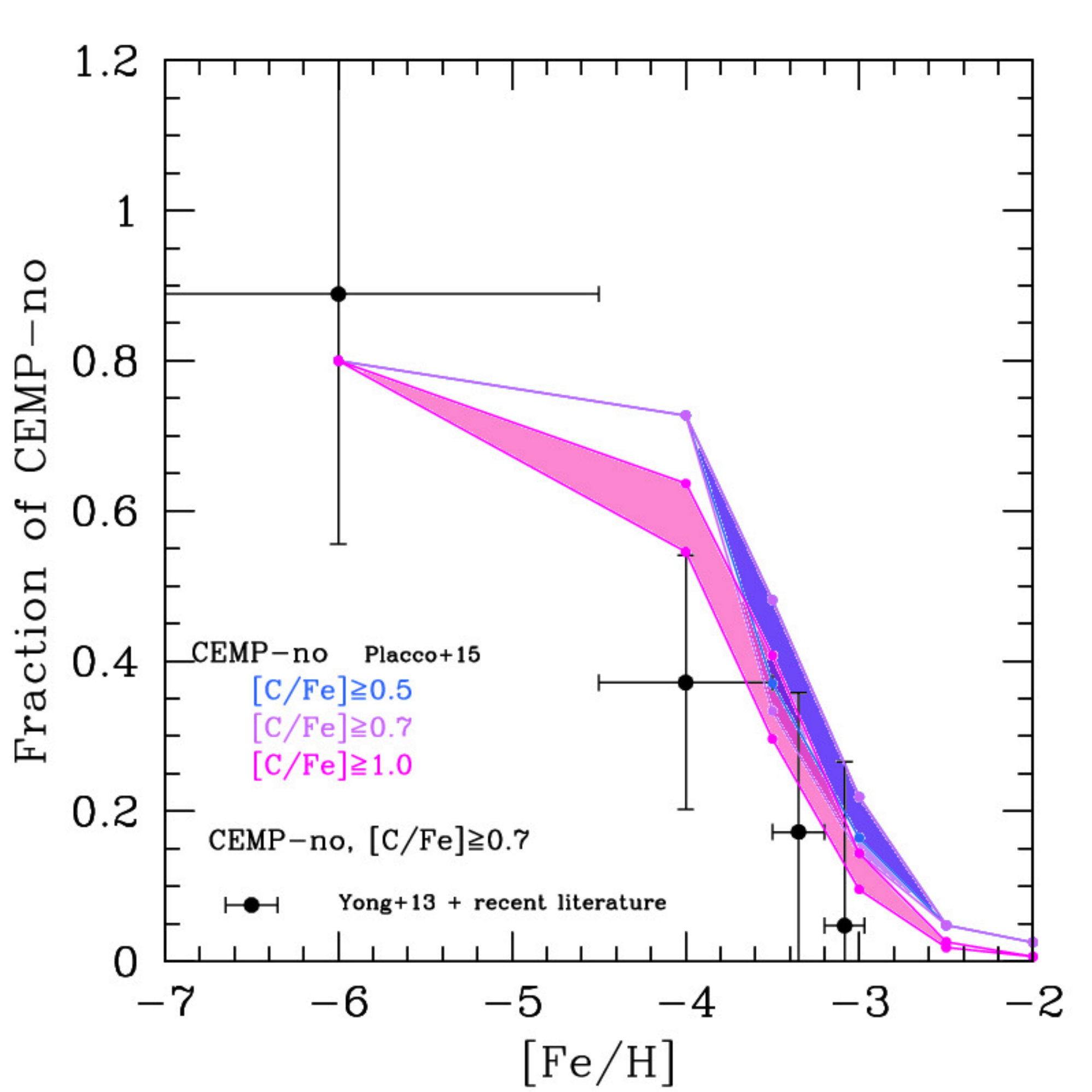}
\caption{Fraction of CEMP-no stars vs [Fe/H] obtained
using different data and [C/Fe] cuts for CEMP-no stars 
(see the labels). We show results from: high-resolution sample by 
\cite{yong13} that we completed by adding more recent literature data
({\it points with poissonian errorbars}); the larger high/medium-resolution 
sample by \cite{placco13} ({\it connected colored points}) that the 
authors corrected to account for carbon depletion due to internal mixing 
processes ({\it upper points}). The shaded area quantifies such a correction.} 
\label{fig:FCEMPref}
\end{figure}
%%%%%%%%%%%%%%%%%%%%%%%%%%%%%%%%%%%%%%%%%%%%%%%%%%%%%%%%%

In Fig.~\ref{fig:MDF}, we show the most recent determinations of 
the low-Fe tail of the Galactic halo MDF as derived by various  
groups, which exploited different data-sets. By normalizing the
MDFs to the same cumulative number of stars at [Fe/H]$\, \leq-3$ we 
compare the results from:
(i) the joint HK and HES medium-resolution surveys, which provide 
$\approx 2765$ stars at [Fe/H]$\, <-2$ \citep{beers05}; 
(ii) the HES survey, which collected $\approx 1500$ stars at 
[Fe/H]$\, <-2$ \citep[e.g.][]{christlieb08}; the derived MDF has been 
corrected by \cite{schoerck09} to account for observational biases 
and incompleteness; 
(iii) the high-resolution sample by \cite{yong13}, who collected 
an homogeneous ensemble of $\approx 95$ stars at [Fe/H]$\, \leq -2.97$ 
by combining data from the literature along with program stars.
Following \cite{schoerck09}, Yong and collaborators corrected 
the high-resolution MDF by using the HES completeness function.  
Furthermore, they accounted for the gaussian error associated 
to each [Fe/H] measurements, and derived a realistically smoothed 
MDF (see their Sec.~3.2 and Fig.~\ref{fig:MDF}). The effects 
of these corrections can be appreciated by comparing the final
high-resolution (HR) MDF in Fig.~\ref{fig:MDF} (shaded area), and 
the points with error bars within $-4\leq \, $[Fe/H]$\, \leq-3$, which 
represent the raw data.\\

By inspecting Fig.~\ref{fig:MDF} we can see that the MDFs obtained 
with the high-resolution sample by \cite{yong13} and the joint HK and HES samples 
by \cite{beers05} are in excellent agreement. Both MDFs continuosly 
decrease between [Fe/H]$\, \approx -3$ and [Fe/H]$\, \approx -4$ spanning 
roughly one order of magnitude in $N_*$. On the other hand, the HES
sample underestimates the total number of stars at [Fe/H]$\, \leq -3.7$. 
As discussed by \cite{yong13}, this is likely due to the selection 
criteria exploited by the HES sample, which reject stars with strong 
G-bands. As high-resolution observations provide more precise [Fe/H] 
measurements, we use the small HR sample by Yong and collaborators
as our ``reference MDF'' for $-4<\,$[Fe/H]$\,<-3$. At lower [Fe/H], we
complete the sample by adding all the stars that have been discovered 
during the years and followed-up at high-resolution (see points in
Fig.~\ref{fig:MDF}). Thus, the reference MDF we compare our
models with rapidly declines with decreasing [Fe/H], exhibits a sharp 
cut-off at [Fe/H]$= -4.2\pm0.2$ and a low-Fe tail made by 9 stars 
that extends down to [Fe/H]$\approx -7.2$ \citep{keller14}.\\ 

Interestingly, 8 out of the 9 Galactic halo stars identified at 
[Fe/H]$\,<-4.5$ show high overabundance of carbon, [C/Fe]$>0.7$
(\citealt{christlieb02,frebel05,norris07,keller14,
hansen15,bonifacio15,allende-prieto15,frebel15,melendez16}, see 
for example Fig.~1 from \citealt{salvadori15}). Although
only 5 of these CEMP stars have available measurements of slow- and 
rapid-neutron capture process elements (s-/r-), which have sub-solar 
values, all of them can be likely classified as ``CEMP-no'' stars 
(e.g. see discussion in \citealt{bonifacio15} and \citealt{norris10}). 
CEMP-no stars do not show s-process elements that are produced by 
AGB stars \citep[e.g.][]{beers05}, they 
are not preferentially associated to binary systems \citep[e.g.][]{hansen13}, 
and they most likely appear at low [Fe/H] (see Fig. 
\ref{fig:FCEMPref}). For these reasons, the chemical abundances 
measured in their photo-spheres are believed to reflect their 
environment of formation, which was likely polluted by Pop~III 
stars that developed mixing and fallback evolving as ``faint 
SNe'' \citep[e.g.][]{bonifacio03,iwamoto05,marassi14}, or by primordial 
``spinstars'', which experienced mixing and mass-loss because 
of high rotational velocities \citep[e.g.][]{meynet06,maeder15}.\\

%%%%%%%%%%%%%%%%%%%%%%%%%%%%%%%%%%%%%%%%%%%%%%%%%%%%%%%%%
\begin{figure*}
\includegraphics [trim=0.5cm 5.5cm 0cm 4cm,
clip=true,width=18.cm]{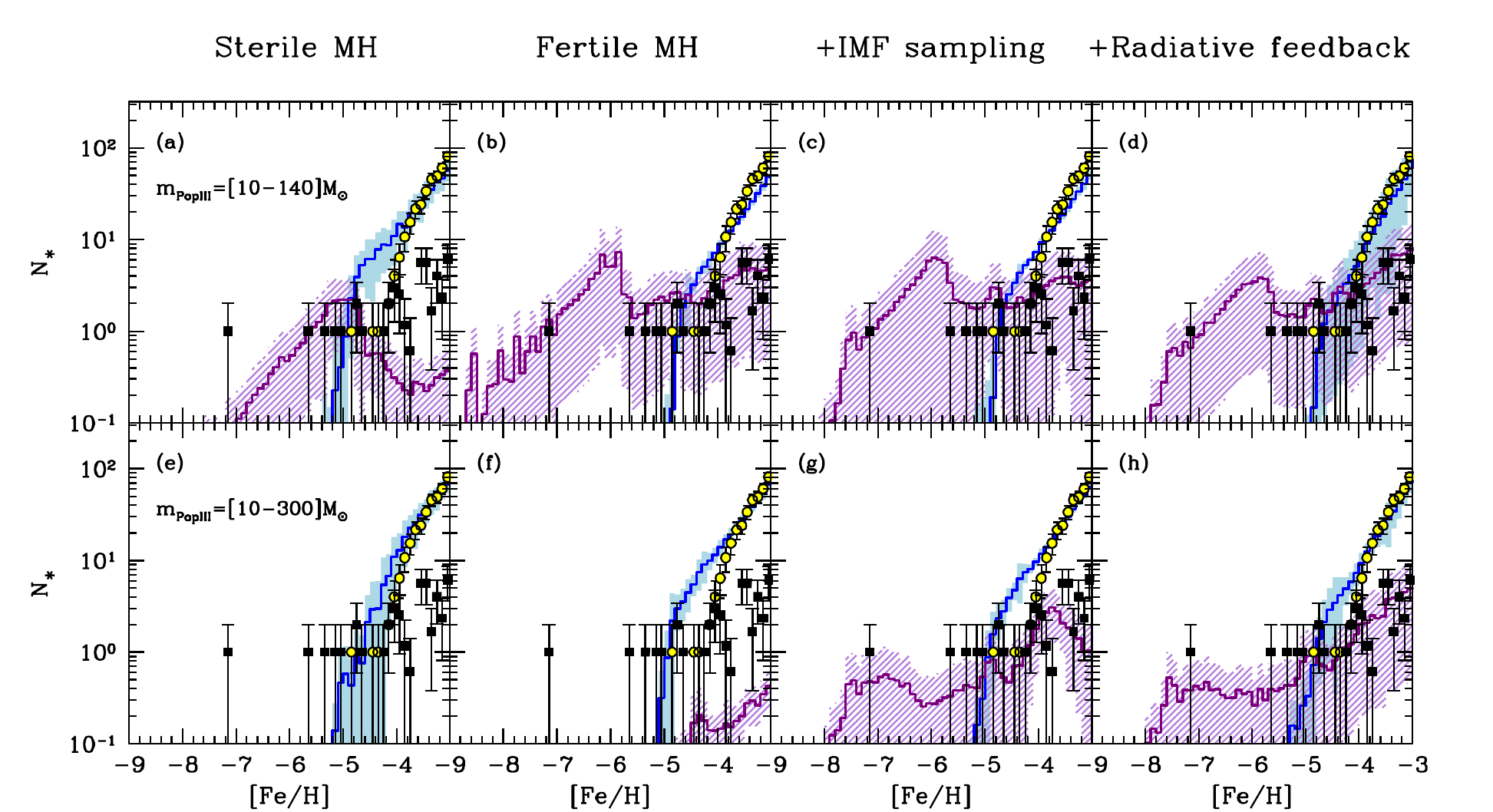}
\caption{Comparison between observed (points with poissonian errorbars) 
and simulated (histograms with shaded areas) Galactic halo MDFs, where we 
differentiate the contribution of C-enhanced (purple histograms with striped-dashed
shaded areas and black filled squares) and 
C-normal (blue histograms with filled shaded areas and yellow filled circles) stars. Pop III stars are assumed to form 
either in the range [10-140] $\Msun $ ({\it top panels}) or [10-300] $\Msun$ 
({\it bottom panels}, see also labels). From left to right we show the results
for models with: sterile mini-haloes ({\it panels a-e}); fertile mini-haloes 
with a temperature-regulated star-formation efficiency ({\it panels b-f}); 
fertile mini-haloes with a temperature-regulated star-formation efficiency and 
a stochastically sampled Pop III IMF ({\it panels c-g}); fertile mini-haloes 
with a star-formation efficiency regulated by radiative feedback and a 
stochastically sampled Pop III IMF ({\it panels d-h}).
}
\label{fig:MDFcases}
\end{figure*}
%%%%%%%%%%%%%%%%%%%%%%%%%%%%%%%%%%%%%%%%%%%%%%%%%%%%%%%%%

In Fig.~\ref{fig:FCEMPref} we show the fraction of CEMP-no stars 
in different [Fe/H] bins, $F_{\tt CEMP-no}=N_{\tt CEMP-no}({\rm [Fe/H]})/N_*({\rm [Fe/H]})$, 
derived by using the HR sample by \cite{yong13} along with new 
literature data. Following \cite{yong13}, we select the iron 
bins in order to have a roughly equal number of stars in each
of them, $N_*\approx 30$. Only in the lowest bin, [Fe/H]$<-4.5$, the 
total number of stars is limited to $N_*=9$. This is reflected in the
larger Poissonian errors. In Fig.~\ref{fig:FCEMPref} we also show the 
$F_{\tt CEMP-no}$ values obtained by using the available data and CEMP-no 
classification by \cite{placco14}. These authors exploited a larger 
sample of high- and medium-resolution observations. Furthermore, they 
corrected the carbon measurements in Red Giant Branch (RGB) stars in 
order to account for the depletion of the surface carbon-abundance, 
which is expected to occur during the RGB phase. We can see that in
both samples $F_{\tt CEMP-no}$ rapidly decreases with increasing [Fe/H],
and that the results are consistent within the Poissonian errors.
However, as already noticed by Frebel \& Norris (2016), the estimated
CEMP-no fraction is higher in the sample by \cite{placco14}. To be
consistent with our reference MDF and limit our comparison to
high-resolution data only, we will focus on the $F_{\rm CEMP-no}$ values
derived by using \cite{yong13} and new literature data. However we
should keep in mind that these are likely lower limits on the actual
CEMP-no fraction at [Fe/H]$\, >-4.5$.

%%%%%%%%%%%%%%%%%%%%%%%%%%%%%%%%%%%%%%%%%%%%%%%%%%%%%%%%%
\section{Results}
\label{sec:res}

In this section we present the results of our models by 
focusing on the MDF and on the properties of CEMP-no stars 
at [Fe/H]$\, <-3$. Since we did not model mass transfer in 
binary systems, we will completely neglect CEMP-s/(rs) stars 
in our discussion \citep[see also][]{salvadori15}.

The comparison between simulated and observed MDFs is shown 
in Fig.~\ref{fig:MDFcases} where the two functions have
been normalized to the same number of observed stars at [Fe/H]
$\, < \, $-3 (see Sec.~\ref{sec:obs}). We computed the mean 
theoretical MDF as the average number of stars in each [Fe/H] 
bin across the 50 merger histories. The 1-$\sigma$ dispersion, 
or standard deviation, is then derived as the square root of 
the variance among the 50 merger histories. Once these quantities 
are evaluated the normalization is done by rescaling the total
number of simulated star at [Fe/H]$\, < \, $-3 to the observed 
value. Note that the same results are obtained by first normalizing 
the simulated MDF of each realization, and then computing the average 
MDF and 1-$\sigma$ dispersion.

By inspecting all the panels, from left to 
right, we can see how the different physical processes included 
in the models affect the shape of the MDF, along with the predicted 
number of CEMP-no stars. For each model, the figure shows how 
the results vary when we assume Pop~III stars to form with 
masses in the range $[10 - 140] \, \Msun$ (top panels), where only faint SNe 
can contribute to metal enrichment, and 
$[10 - 300]\, \Msun$ (bottom panels), where both faint SNe and 
PISNe can produce and release heavy elements. 

Panels (a, e) show the results for models with 
``sterile'' mini-haloes, i.e. assuming that systems with 
$T_{\rm vir} \, < \, 10^4$~K are not able to form stars.  
As discussed in \cite{debennassuti14}, we see that both the 
low-Fe tail of the MDF, at [Fe/H]$\, < \, -5$, and the number 
of CEMP-no stars, can only be reproduced when the Pop~III 
stars have masses $m_{\rm popIII} = [10-140] \, \Msun$ (panel a).
This is because CEMP-no stars form in environments enriched by 
the chemical products of primordial faint SNe, which produce 
large amount of C and very low Fe. These key chemical signatures 
are completely washed out if PISNe pollute the same environments (panel e), 
because of the large amount of Fe and other heavy elements that
these massive stars produce ($\approx 50\%$ of their masses). 
Panel (a) also shows that, although the overall 
trend is well reproduced, the model predicts an excess of 
C-normal stars at [Fe/H]$ \, < \, -4$, and it underestimates the 
observations at [Fe/H]$ \, < \, -6$. In other words, the slope of
the total MDF (blue+purple histograms) is predicted to be 
roughly constant, at odds with observations (see also Sec.~\ref{sec:obs} 
and the discussion in \citealt{debennassuti14}).

The results change when mini-haloes are assumed to be ``fertile'' 
(panels b, f) and their star formation efficiency is assumed to be 
simply regulated by their virial temperature (see Sec.~\ref{sec:cool}).
Note that the evolution of the minimum halo mass for star-formation 
assumed in this case \citep{salvadori09,salvadori12} is consistent 
with the mass threshold for efficient H$_2$-cooling obtained by adopting
the LW background of \cite{ahn09}, as it has been shown by \cite{salvadori14}.

Panel (b) shows that the overall MDF has a different shape with 
respect to panel (a): it rapidly declines around [Fe/H] $\, 
\approx -4.5$ and shows a low-Fe tail that extends down to 
[Fe/H]$\, \approx -8$ and that is made by CEMP-no stars {\it only}. 
Such a discontinuous shape, which is consistent with observations, 
reflects the different environment of formation of CEMP-no and 
C-normal stars (see also the discussion in \citealt{salvadori16}). 
CEMP-no stars are formed in low-mass mini-haloes at a very low and almost 
constant rate (e.g. \citealt{salvadori15}). C-normal stars, instead, 
predominantly form in more massive systems, which more efficiently 
convert gas into stars, producing the rapid rise of the MDF at [Fe/H]$\, > \, -5$.

The comparison between panel (b) and (c) shows that when
$m_{\rm popIII}=[10-140]\, \Msun$ the MDF does not depend on
the incomplete sampling of the Pop~III IMF (see Sec.~\ref{sec:stocimf}). 
However, the picture changes when $m_{\rm popIII}=[10-300]\,\Msun$ 
(panel g). Because of the poor sampling of the Pop~III IMF in mini-haloes, 
the formation of stars in the faint SN progenitor mass range is strongly 
favored with respect to more massive PISN progenitors (Fig.~\ref{fig:imf}).  
As a consequence, the chemical signature of primordial faint SNe is 
retained in most mini-haloes, where CEMP-no stars at [Fe/H]$\, < \, -4$ 
preferentially form. Furthermore, because of the reduced number of Pop~III 
haloes imprinted by primordial faint SN {\it only}, the amplitude of the 
low-Fe tail is lower than in panel (c), and thus in better agreement 
with observations.

Panels (d, h) show the results obtained by self-consistently computing 
the star-formation efficiency in mini-haloes (Sec.~\ref{sec:cool}). 
We note that there are no major changes with respect to panels 
(c, g), meaning that our implementation of the radiative feedback described 
in Sec.~\ref{sec:rad} is consistent with the simple analytical prescriptions 
used in panels (c, g). In other words, our model results do not depend strongly 
on the LW flux (upper panel of Fig.~\ref{fig:radiation}), which in such a 
biased region of the Universe is expected to be larger than in the average 
cosmic volume \citep[e.g.][]{ahn09,xu16}. Still, three important differences 
can be noticed by inspecting panels (d, h): (i) the number 
of C-normal stars shows a larger scatter than previously found,  
(ii) their excess with respect to the data is partially reduced, and
(iii) the number of CEMP-no star with [Fe/H]$\, > \, -4$ is larger, 
and thus in better agreement with available data.
These are consequences of the modulation of mini-haloes star-formation efficiency,
which declines with cosmic time as a consequence of the decreasing mean gas 
density and of the increasing radiative feedback effects by the growing LW 
background (Fig.~\ref{fig:cooling}). This delays metal-enrichment, preserving 
the C-rich signatures of faint SNe in the smallest Lyman-$\alpha$ cooling haloes, 
where CEMP-no stars with [Fe/H]$\, > -4$ can form with higher efficiency.

We finally note that in all models the scatter of the MDF is much 
larger for the CEMP component than for the C-normal one. This is due to the 
broad dispersion among different merger histories at $z>10$ (see Fig.~4 of 
\citealt{debennassuti14}), where the CEMP MDF is built (see their Fig.~10).

When $m_{\rm popIII}=[10-140]\, \Msun$, furthermore, a bump in the 
simulated CEMP MDF is observed. This is due to the fact that in this case Pop~III
stars evolve only as faint SNe, polluting the ISM of their their hosting 
(mini-)haloes around $[Fe/H]_{Pop~III} \approx -6$, a value that is settled 
by the Fe yields of faint SNe and mean star-formation efficiency of mini-haloes 
(for more massive Ly-$\alpha$ cooling haloes $\rm [Fe/H]_{PopIII} \sim -5$).
Once normal Pop~II stars begin to evolve as core-collapse SNe, the Fe enrichment 
is much faster because of the larger Fe yields. Thus, the ISM is enriched up to 
higher [Fe/H], giving origin to the bump of CEMP stars.

\begin{figure}
\centering
\includegraphics[trim=6.5cm 1.cm 6.5cm 1.cm,clip=true,width=0.8\columnwidth]{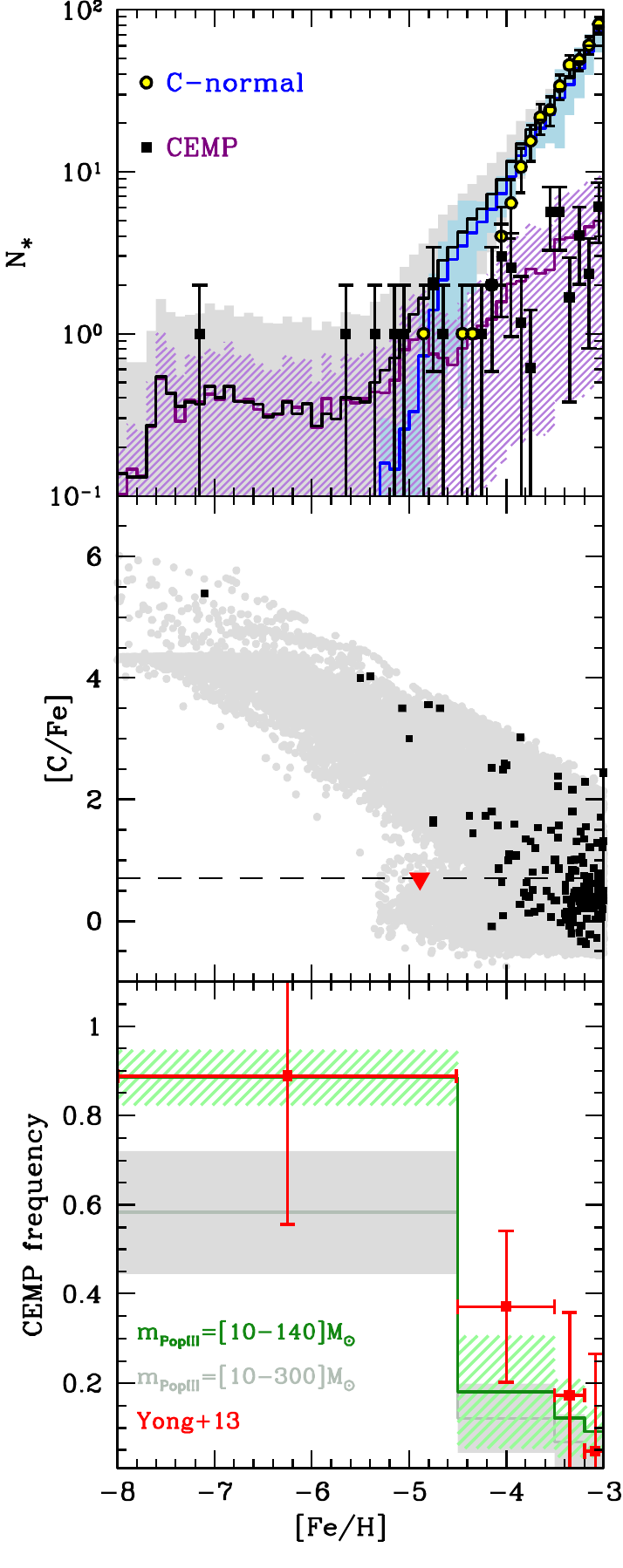}
\caption{{\it Top:} same as panel (h) in Fig.~\ref{fig:MDFcases}, but with the inclusion of the total MDF 
and errors induced by the observations, which have been obtained using a Monte
Carlo selection procedure (see text). {\it Middle:} stellar [C/Fe] vs [Fe/H]
measured in Galactic halo stars (black squares, see Fig.~1 of
\citealt{salvadori15}) and obtained in 50 realizations of our fiducial model
for $m_{\rm popIII}$ = [10-300]\, $\Msun$ (grey dots). The red triangle at [Fe/H]$\sim$-5 shows the upper limit for the only C-normal metal-poor star observed at [Fe/H]$\lesssim$-4.5 so far \citep{caffau11}. The line shows the value of 
[C/Fe]$=0.7$, 
which discriminates between CEMP-no and C-normal stars. {\it Bottom:} comparison 
between the observed fraction of CEMP-no stars (points with Poissonian errorbars as 
explained in Sec.~\ref{sec:obs}) and the fraction predicted by our fiducial models for different 
Pop~III mass ranges: 
$m_{\rm popIII}$ = [10-140]\, $\Msun$ as green histogram with striped-dashed shaded area; 
$m_{\rm popIII}$ = [10-300]\, $\Msun$ as grey histogram with filled shaded area.}
\label{fig:fiducial}
\end{figure}

When we assume $m_{\rm popIII}=[10-300]\, \Msun$, we obtain a better 
agreement with the data at [Fe/H]$>-4$, i.e. where the statistics 
of observed stars is higher. However at lower metallicities, $-5\, <\, 
$[Fe/H]$\, <\, -4$, we can see that both models over-predict the total 
number of stars with respect to current data. But how significant is the 
number of observed stars at these [Fe/H]? In the top panel of 
Fig.~\ref{fig:fiducial} we plot the result of our fiducial model 
for $m_{\rm popIII} = [10-300] \, \Msun$, also including the 
{\it total} MDF and the intrinsic errors induced by observations 
(grey shaded area). These errors are evaluated by using a Monte 
Carlo technique that randomly selects from the theoretical MDF a 
number of stars equal to the observed one \citep[see Sec.~5 of][]{salvadori15}. 
The results, which represent the average value $\pm1-\sigma$
errors among 1000 Monte Carlo samplings, allow us to quantify 
the errors induced by the limited statistics of the observed 
stellar sample. We can see that at 
[Fe/H]$<-4$ these errors are larger than the spread induced by 
different realizations, implying that the statistics should increase before 
drawing any definitive conclusions (see also Sec.~\ref{sec:disc} for a discussion). 
We can then have a look to other observables.

%%%%%%%%%%%%%%%%%%%%%%%%%%%%%%%%%%%%%%%%%%%%%%%%%%%%%%%%%
%%%%%%%%%%%%%%%%%%%%%%%%%%%%%%%%%%%%%%%%%%%%%%%%%%%%%%%%%

%%%%%%%%%%%%%%%%%%%%%%%%%%%%%%%%%%%%%%%%%%%%%%%%%%%%%%%%%%%%%%%%%%%%%%%
\subsection{CEMP and carbon-normal stars}           
\label{sec:cemp}
In the middle panel of Fig.~\ref{fig:fiducial} we compare our model results 
with the available measurements of the [C/Fe] ratio for Galactic halo stars 
at different [Fe/H]. 
Our findings show a decreasing [C/Fe] value for increasing 
[Fe/H], in good agreement with observations.
According to our results, 
stars at [Fe/H]$<-5$ formed in mini-haloes polluted by primordial faint SNe 
(see also Fig.~\ref{fig:2g}). When these stars explode, they release 
large amounts of C and very small of Fe, i.e. [C/Fe]$_{\rm ej}> 4$, 
thus {\it self-enriching} the ISM up to total metallicities $Z>10^{-4}\Zsun$. 
The subsequent stellar generations formed in these mini-haloes are thus
``normal'' Pop~II stars. Their associated core-collapse SNe 
further enrich the ISM with both Fe and 
C, i.e. [C/Fe]$_{\rm ej}\approx 0$, thus producing a gradual decrease of 
[C/Fe] at increasing [Fe/H]. Such a trend is reflected in the chemical
properties of long-lived CEMP-no stars that formed in these environments 
(see middle panel of Fig.~\ref{fig:fiducial}).

When and where do the most metal-poor C-normal stars form? 
We find that at $ \rm [Fe/H] < -4.5$, C-normal stars can only form in 
haloes with a dust-to-gas ratio above the critical value, and that have 
{\it accreted their metals and dust from the surrounding MW environment}. 
When this occurs, normal Pop~II SNe in self-enriched haloes have already 
become the major contributors to the metal enrichment of the external MW 
environment, leading to $\rm [C/Fe] < 0.7$ (see also the middle panel of 
Fig.~7 in \citealt{debennassuti14}).
As star-formation proceeds in these Pop~II haloes, more core-collapse SNe 
contribute to self-enrichment of these environments, thus further increasing 
[Fe/H] with an almost constant [C/Fe]. This creates the horizontal branch 
shown in the middle panel of Fig.~\ref{fig:fiducial} within 
[Fe/H]$\approx[-5,-3]$. In this region of the plot, we note that the model 
predicts a larger concentration of C-normal stars than currently observed.\\

In the bottom panel of Fig.~\ref{fig:fiducial} we compare the expected 
frequency of CEMP-no stars at different [Fe/H] with the available data 
(see Sec.~\ref{sec:obs}). We can see that our models over-estimate the contribution 
of C-normal stars at [Fe/H]$<-4$. Interestingly, a better agreement is 
obtained when the mass of Pop~III stars is limited to the range 
$m_{\rm popIII} = [10 - 140]\, \Msun$, since more faint SNe are produced. 
The number of stars observed at [Fe/H]$<-4$ 
is limited to $\approx 10$ (see Sec.~\ref{sec:obs}). Furthermore, such a discrepancy 
between model and data might also be connected to some underlying physical 
processes that cannot be captured by the model, as we will extensively 
discuss in Sec.~\ref{sec:disc}. In conclusion, Pop~III stars with masses $m_{\rm popIII} 
= [10 - 300] \, \Msun$ are favored by our analysis. Such a model better 
matches the overall shape of the MDF of CEMP-no and C-normal stars, and 
it is also in better agreement with data at [Fe/H]$>-4$ (compare panels 
d and h of Fig.~\ref{fig:MDFcases}). Only at these high [Fe/H] the 
intrinsic observational errors are lower than the dispersion induced 
by different merger histories (see top panel of Fig.~\ref{fig:fiducial}), making the 
difference between model and data statistically significant.

%%%%%%%%%%%%%%%%%%%%%%%%%%%%%%%%%%%%%%%%%%%%%%%%%%%%%%%%%
\subsection{Predictions for second-generation stars}           
\label{sec:2G}

%%%%%%%%%%%%%%%%%%%%%%%%%%%%%%%%%%%%%%%%%%%%%%%%%%%%%%%%%
\begin{figure}
\centering
\includegraphics[trim= 0.5cm 1.5cm 1.5cm 1.5cm,clip=true,width=1.\columnwidth]{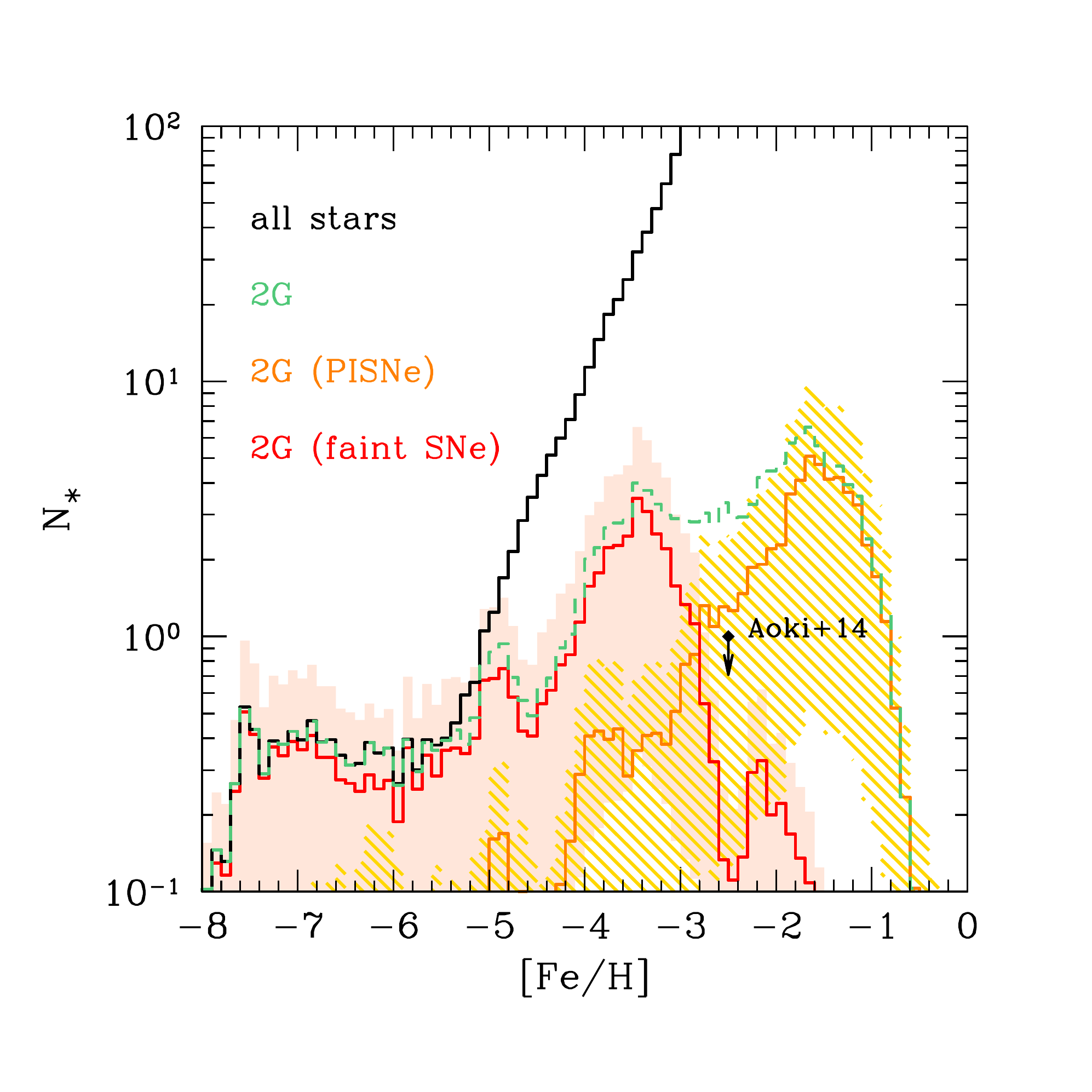}
\caption{Simulated average MDF for all stars (black histogram) and for 2G stars (dashed green histogram), selecting them as stars formed in environments where metals come mostly ($>$ 50\%) from Pop~III stars. The red (orange) histogram represents 2G stars with a dominant ($>$ 50\%) metal contribution from Pop~III faint SNe (PISNe) and the filled (striped-dashed) region represents the 1-$\sigma$ dispersion. The upper limit 
at [Fe/H] = -2.5 represents the recent observation of a 
star with chemical imprint of PISNe \citet{aoki14}.}
\label{fig:2g}
\end{figure}
%%%%%%%%%%%%%%%%%%%%%%%%%%%%%%%%%%%%%%%%%%%%%%%%%%%%%%%%%
Our analysis of the Galactic halo MDF and properties of CEMP-no vs C-normal stars supports 
the most recent findings of numerical simulations for the formation of the first stars, which indicate
that Pop~III stars likely had masses in the range $m_{\rm popIII} = [10-300] \, \Msun$. 
We make one step further, and we quantify the expected number and typical [Fe/H]
values of second-generation (2G) stars, i.e. stars that formed in gaseous environments that
were predominantly polluted by Pop~III stars \citep{salvadori07}.
%%%%%%%%%%%%%%%%%%%%%%%%%%%%%%%%%%%%%%%%%%%%%%%%%%%%%%%%%%%%%%%%%%%%%%%%%%%
\begin{figure} 
\centering
\includegraphics [trim=1.9cm 2cm 9.cm 2.cm, clip=true,width=0.8\columnwidth]{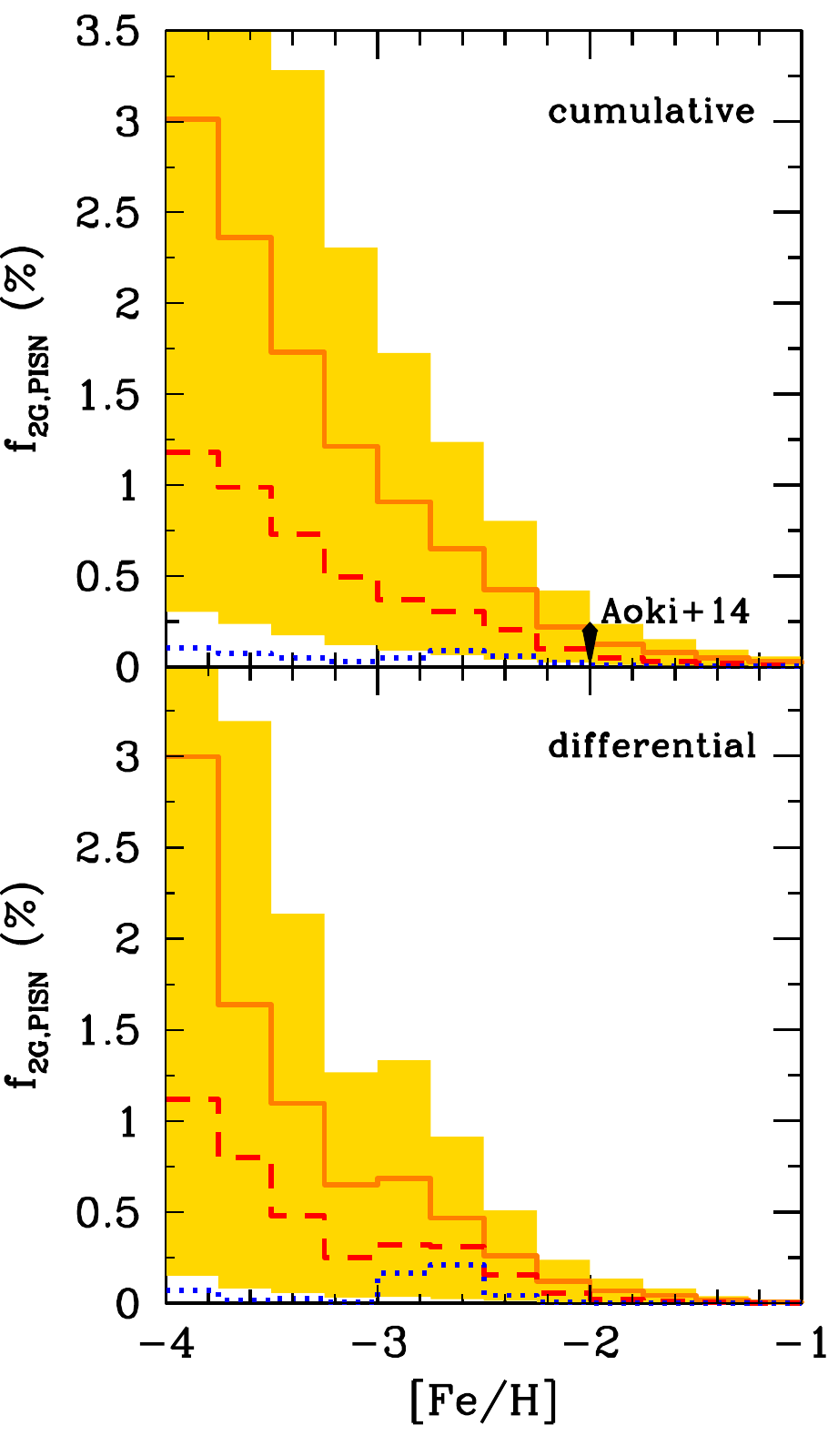}
\caption{Cumulative (top) and differential (bottom) fraction ($\%$) 
of 2G stars imprinted by PISNe with respect 
to the total number of stars in different [Fe/H] bins. The colors 
show the percentage of 2G stars formed in environments where 
metals from PISNe correspond to at least 50\% (solid orange), 80\% (dashed 
red) and 99\% (dotted blue) of the total. The yellow shaded area is 
the 1-$\sigma$ dispersion for the 50\% case. The data point is from 
\citet{aoki14}, who possibly detected the chemical imprint 
of PISNe (upper limit) in 1 out of 500 stars at [Fe/H]$<-2$.}
\label{fig:2gperc}
\end{figure}
%%%%%%%%%%%%%%%%%%%%%%%%%%%%%%%%%%%%%%%%%%%%%%%%%%%%%%%%%%%%%%%%%%%%%%%%%%%

In Fig.~\ref{fig:2g} we show the predictions for our fiducial model with $m_{\rm popIII} = [10-300]\, \Msun$.
The figure illustrates the total distribution of 2G stars formed in environments in which Pop~III stars
provided $\geq 50\%$ of the total amount of metals (green). Furthermore, it shows the individual
distributions for 2G stars imprinted by faint SNe (red) and PISNe (orange). It is clear that the total
number of 2G stars represents an extremely small fraction of the Galactic halo population, in 
full agreement with previous studies \citep{salvadori07,karlsson08}. Furthermore, we can see that 
the distributions of 2G stars polluted by faint SNe and PISNe are extremely different. Because of the 
low amount of Fe released by faint SNe, 2G stars enriched by this stellar population predominantly 
appear at [Fe/H]$<-3$. Interestingly, we find that their distribution essentially matches that of
CEMP-no stars (see Fig.~\ref{fig:fiducial}), representing $\approx 100\%$ of the observed stellar population 
at [Fe/H]$<-5$. In other words, {\it our model predicts that all CEMP-no stars in this [Fe/H] 
range have been partially imprinted by Pop~III faint SNe}. On the contrary, we see that the 
distribution of 2G stars polluted by PISNe is shifted towards higher [Fe/H] values. This is 
because massive PISNe produce larger amount of iron than faint SNe, thus self-enriching their 
birth environments up to [Fe/H]$>-4$.
In fact, for $m_{\rm popIII} = [10-300]\, \Msun$, the Fe yielded 
by PISN per total stellar mass formed, $M_*$, is $\rm Y^{PISN}_{Fe}\approx 
M_{Fe}/M_\ast \approx 2.7\times10^{-2}$, which is several orders of magnitude
larger than the one produced by faint SNe. As described in Sec.~\ref{sec:stocimf} 
a mini-halo can likely host a PISN if it forms a stellar mass $\rm M_\ast \gtrsim 10^4 
\Msun$ which implies a Fe production of $\rm M_{Fe} \gtrsim 270 \Msun$. According 
to our prescription for star-formation, the typical mass of such a mini-halo 
is $\rm \sim 10^8 \Msun$, meaning that this amount of Fe can be dispersed in a
gaseous environment of $\sim 10^7 \Msun$, assuming the cosmological baryon 
fraction. This leads to an ISM polluted at [Fe/H]$\sim$-1.8, which is consistent 
with the peak of 2G stars with PISN imprint predicted by our model (see Fig.~\ref{fig:2g}).
Unfortunately, at these [Fe/H] the Galactic halo population
is dominated by stars formed in environments mostly polluted by normal Pop II SNe 
(see Sec.~\ref{sec:cemp}). This makes the detection of 2G stars imprinted by PISNe 
very challenging.

In the upper (lower) panel of Fig.~\ref{fig:2gperc} we quantify the cumulative (differential) fraction of 2G stars 
imprinted by PISNe with respect to the overall stellar population as a function of [Fe/H]. In 
both panels we identify 2G stars that formed in environments where 50\%, 80\%, and 99\% of 
the metals were coming from PISNe. We can clearly see that
 2G stars always represent $<3\%$ of the total stellar population. 
Furthermore, the cumulative fraction of 2G stars polluted by PISN at 50\% (80\%) level, strongly 
decreases with increasing [Fe/H]: around [Fe/H]$= -2$, in particular, these 2G stars are predicted
to represent 0.25\% (0.1\%) of the total stellar population, which is fully consistent with 
current observations (see point in Fig.~\ref{fig:2gperc}). Indeed, among the $\approx 500$ Galactic halo stars 
analyzed so far at [Fe/H]$<-2$, there is only one candidate at [Fe/H]$\approx-2.4$ that might
have been imprinted by a PISN \citep{aoki14}. This rare Galactic halo star shows several peculiar 
chemical elements in its photo-sphere, which might reflect an ISM of formation enriched by both
massive PISNe and normal core-collapse SNe \citep{aoki14}. In Fig.~\ref{fig:2gperc} we 
can see that 2G stars imprinted by the chemical products of PISNe {\it  only}, represent $<0.1\%$ 
of the total Galactic halo population (top), and they are predicted to be more frequent at 
$-3<$[Fe/H]$<-2$ (bottom).\\ 

%%%%%%%%%%%%%%%%%%%%%%%%%%%%%%%%%%%%%%%%%%%%%%%%%%%%%%%%%
\subsection{Varying the Pop~III IMF}
\label{sec:popIII_IMF}
\begin{figure*}
\includegraphics [trim=0cm 7cm 0cm 6cm, clip=true,width=18.cm]{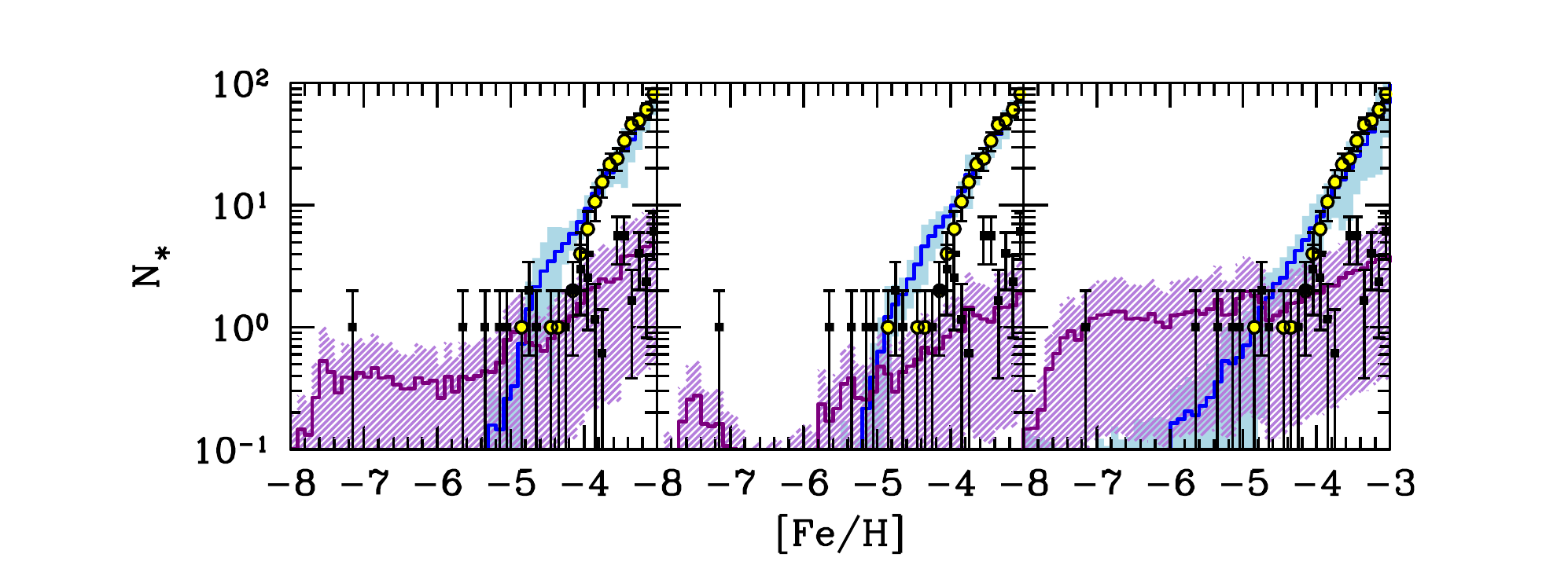}
\caption{Comparison between the observed and simulated Galactic halo MDFs (see Fig.~\ref
{fig:MDFcases}) obtained by using different IMF for Pop~III stars: a Larson IMF 
with $m_{popIII}$=[10-300]$\Msun$ and $m_{\rm ch} = 20 \Msun$ ({\it left}), a 
Flat IMF with $m_{popIII}$=[10-300)]$\Msun$ ({\it middle}), and a Larson IMF 
with $m_{popIII}$=[0.1-300]$\Msun$ and $m_{\rm ch} = 0.35 \Msun$ ({\it right}).}
\label{fig:MDF_IMF}
\end{figure*}

%%%%%%%%%%%%%%%%%%%%%%%%%%%%%%%%%%%%%%%%%%%%%%%%%%%%%%%%%%%%%%%%%%%%%%%%%%%%
We can finally analyze the dependence of our findings on the slope
and mass range of the Pop~III IMF. These results are shown in Fig.~\ref{fig:MDF_IMF}, 
where we compare our reference model (left), with 
two alternative choices of the Pop~III IMF, inspired by numerical 
simulations: a flat IMF with $m_{\rm popIII} = [10 - 300] \, \Msun$ 
\citep[e.g.][]{hirano14}, and a Larson IMF with $m_{\rm popIII} = [0.1 - 300]\,
\Msun$ and a characteristic mass $m_{\rm ch} = 0.35 \, \Msun$ \citep[e.g.][]{stacy16}.

The low-Fe tail of the MDF, and thus the number and distribution 
of CEMP-no stars, strongly depends on the  
Pop~III IMF. When a flat IMF is considered (middle panel of Fig.~\ref{fig:MDF_IMF}), 
less CEMP-no stars are produced with respect 
to the reference case (left panel), 
both at low and at high [Fe/H]. This is due to the smaller 
fraction of faint SNe with respect to the total mass of Pop~III 
stars formed. This is equal to $\approx 1.5\%$ for a 
flat IMF, and $\approx 70\%$ for a Larson IMF, where these numbers
have been obtained by integrating the normalized IMFs in the mass range $[8 - 40]\, \Msun$.
Such a large difference 
is partially mitigated by the incomplete sampling of the Pop~III IMF 
in inefficient star-forming mini-haloes (see Sec.~\ref{sec:stocimf}). 
Still, the 
different IMF slope decreases substantially the number of CEMP-no stars, 
making the flat IMF model {\it partially inconsistent} with current data-sets. 

%%%%%%%%%%%%%%%%%%%%%%%%%%%%%%%%%%%%%%%%%%%%%%%%%%%%%%%%%%%%%%%%%%%%%%%%%%%%
%%%%%%%%%%%%%%%%%%%%%%%%%%%%%%%%%%%%%%%%%%%%%%%%%%%%%%%%%%%%%%%%%%%%%%%%%%%%

On the other hand, when the Pop~III IMF is extended down to lower
masses, $m_{\rm popIII} = [0.1 - 300] \,\Msun$, the number of CEMP stars 
increases with respect to the reference model (right vs left panels 
of Fig~\ref{fig:MDF_IMF}). However, if we make the same calculation 
as before, we find faint SNe to represent 
$\approx 1\%$ of the total Pop~III stellar mass, substantially less than
in our reference model. The larger number of CEMP stars at low [Fe/H]
must then have a different origin.

%According to our predictions $\approx 30\%$ of CEMP stars with [Fe/H] $ < -5$ form out of gas that 
%has been polluted by the winds of Pop~III Asymptotic Giant Branch (AGB) stars, 
%with $m_{\rm popIII} = [2 - 8]\,\Msun$. Thus, these CEMP stars should also 
%retain the peculiar chemical signature of $Z = 0$  AGB stars, 
%such as the slow-neutron capture elements \citep[e.g.][]{goriely01,
%goriely04}. In other words, we predict $\approx 30\%$ of them to be CEMP-s 
%stars rather than CEMP-no stars. This seems to be in contrast with 
%current observational findings for CEMP stars at [Fe/H]$<-3$ 
%\citep[e.g.][]{norris10,bonifacio15}, although a larger statistics 
%is required to rule out this Pop~III IMF model.

At [Fe/H] $< -5$, we predict that $\approx 35\%$
\footnote{This is an average value for $-8 <$[Fe/H]$< -5$. 
Within this [Fe/H] range the percentages can vary from 
$\approx$ 20 to $\approx$ 60 $\%$.} of second-generation CEMP stars
($\approx 15\%$ of the total number of stars)
have been imprinted by a mixed population of Pop~III stars, including
Asymptotic Giant Branch (AGB) stars with $m_{\rm popIII} =[2 - 10]\,\Msun$. 
In particular we find that these Pop~III AGB stars might provide from 
$\approx$ 10 to $\approx$ 30 $\%$ of the total amount of heavy elements polluting the birth 
environment of these 2G stars. These CEMP stars, therefore, might also retain 
the peculiar chemical signature of Pop~III AGB stars, such as the slow-neutron 
capture elements \citep[e.g.][]{goriely01,goriely04}. In other words $\approx 35\%$ 
of these [Fe/H]$< -5$ stars might be CEMP-s stars rather than CEMP-no stars.
This seems to be in contrast with current observational findings \citep[e.g.][]{norris10,
bonifacio15} although more work needs to be done to rule out this Pop~III IMF 
model. On the one hand detailed theoretical calculations are required to understand 
if Pop~III AGB stars produce s-process elements at observable levels.
On the other hand, a larger stellar sample at [Fe/H] $< -5$ is mandatory 
to robustly constrain the low-mass end of the Pop III IMF.

Furthermore the model predicts the existence of metal-free stars, as 
long as stars with $Z<10^{-5}\, \Zsun$, that should respectively correspond 
to $\approx 0.15\%$ and $\approx 0.3\%$ of the total number of stars at
[Fe/H]$<-3$.
As already discussed and 
quantified by several authors \citep{tumlinson06,salvadori07,hartwig15}, 
this result might be in contrast with the current {\it non-detection} of 
metal-free stars. Yet, given the total number of stars collected 
at [Fe/H]$<-3$, which is $\approx 200$, the fraction of zero-metallicity 
stars should be $<1/200=0.5\%$, which is still consistent with our findings.

%%%%%%%%%%%%%%%%%%%%%%%%%%%%%%%%%%%%%%%%%%%%%%%%%%%%%%%%%%%%%%%%%%%%%
\section{Summary and discussion}
\label{sec:disc}
%%%%%%%%%%%%%%%%%%%%%%%%%%%%%%%%%%%%%%%%%%%%%%%%%%%%%%%%%%%%%%%%%%%%%%
In this paper, we investigate the role that Pop~III star-forming 
mini-haloes play in shaping the properties of the Metallicity 
Distribution Function (MDF) of Galactic halo stars and the relative 
fraction of CEMP-no and C-normal stars observed at [Fe/H]$<-3$. 
To this end, we use the merger tree code \textsc{gamete} \citep{salvadori07,salvadori08}, 
which we further implement with respect to recent studies for Galactic 
halo stars \citep{debennassuti14} to resolve H$_2$-cooling mini-haloes 
with $T_{\rm vir}< 10^4$~K, which are predicted to host the first, Pop~III 
stars \citep[e.g.][]{abel02,hirano14}. Following \cite{debennassuti14}, 
we initially assumed Pop~III stars to have masses in the range $[10 - 300]\,\Msun$,
and to be distributed according to a Larson IMF (Fig.~\ref{fig:imf}). 
We subsequently explored the dependence of our results on different 
IMF slope and Pop~III mass range. In all cases, we assumed that 
Pop~III stars with masses $m_{\rm popIII} = [10 - 40]\, \Msun$ evolved 
as faint SNe that experience mixing and fallback 
\citep[e.g.][]{bonifacio03,umeda03,iwamoto05,marassi14,marassi15}.

To accurately model the formation of Pop~III stars, we introduce a 
new random IMF selection procedure, which allows us to account for the 
incomplete sampling of the Pop~III IMF in inefficiently star-forming 
mini-haloes. To compute the star-formation efficiency of these 
low-mass systems, we exploit the results of numerical simulations 
by \cite{valiante16}, who evaluate the cooling properties of H$_2$-cooling 
mini-haloes as a function of: $(i)$ virial temperature, $(ii)$ formation redshift, 
$(iii)$ metallicity, and $(iv)$ Lyman Werner (LW) background. To this end, we 
self-consistently compute the LW and ionizing  photon fluxes produced 
by Milky Way (MW) progenitors, along with the reionization history 
of the MW environment, which is consistent with recent theoretical findings 
\citep{salvadori14} and new observational data \citep{planck2015,planck2016}.\\

The main results of our work, and their implications for theoretical 
and observational studies, can be summarized as follows:
\begin{itemize}  
\item     
the shape of the low-Fe tail of the Galactic halo MDF is correctly 
reproduced only by accounting for star-formation in mini-haloes, 
which confirms their key role in the early 
phases of galaxy formation. 
\item     
We demonstrate that it is fundamental to account for the poor sampling of the 
Pop~III IMF in mini-haloes, where inefficient Pop~III starbursts, with
$<10^{-3} \, \Msun \rm \, yr^{-1}$, naturally limit the formation of $>100 \, \Msun$ 
stars  and hence change the ``effective'' Pop~III IMF.
\item
CEMP-no stars observed at [Fe/H]$<-3$ are found to be imprinted by 
the chemical products of primordial faint SNe, which provide $>50\%$ 
of the heavy elements polluting their birth environment, making them 
``second-generation'' stars
\item
Second-generation stars imprinted by PISNe, instead, emerge at $-4<$
[Fe/H]$<-1$, where they only represent a few $\%$ of the total 
halo population, which makes their detection very challenging.
\item
At [Fe/H]$\approx -2$, only $0.25\%$ ($0.1\%$) of Galactic halo
stars are expected to be imprinted by PISNe at $>50\%$ ($>80\%$) level, 
in good agreement with current observations.
\item     
The low-Fe tail of the Galactic halo MDF and the properties of CEMP-no 
stars strongly depends on the IMF shape and mass range of Pop~III stars.
\end{itemize}   

A direct implication of our study is that the Galactic halo MDF is a key 
observational tool not only to constrain metal-enrichment models of MW-like 
galaxies and the properties of the first stars \citep{tumlinson06,salvadori07}, 
but also the star-formation efficiency of mini-haloes. These observations, 
therefore, can be used to complement data-constrained studies of ultra-faint 
dwarf galaxies aimed at understanding the properties of the first star-forming 
systems \citep[e.g.][]{salvadori09,bovill09,blandhawthorn15,salvadori15}. 

A key prediction of our model concerns the properties and frequency of 
second-generation (2G) stars formed in gaseous environments imprinted 
by $>50\%$ of heavy elements from PISNe. In agreement with previous 
studies \citep{salvadori07,karlsson08}, we find that these 2G stars are 
extremely rare, and we show that their MDF peaks around [Fe/H]$= -1.5$ 
(Fig.~\ref{fig:2g}). In particular, 2G stars imprinted by PISNe {\it only} ($>99\%$ 
level) are predicted to be more frequent at $-3<$[Fe/H]$<-2$, where 
they represent $\approx 0.1\%$ of the total. In the same 
[Fe/H] range, 2G stars polluted by PISNe at $>50\%$ ($>80\%$) level, 
constitute $\approx 0.4\%$ ($\approx 0.2\%$) of the stellar population. 
These numbers are consistent with the unique detection of a rare halo 
star at [Fe/H]$\approx -2.5$ that has been possibly imprinted {\it also} 
by the chemical products of PISNe \citep{aoki14}.

On the other hand, we show that {\it C-enhanced stars at [Fe/H] $< -5$ are
all truly second generation stars}. Hence, the number and properties of these CEMP stars
can provide key indications on the Pop~III IMF. 

%Given the current statistics, we show that a flat Pop III IMF with $m_{\rm popIII} = [10 - 300]\, \Msun$
%is disfavoured by the observations. Furthermore, by assuming a 
%Larson IMF with $m_{\rm popIII} = [0.1 - 300]\, \Msun$ and $m_{\rm ch} = 0.35 \, \Msun$, we find 
%that $\approx 30 \%$ of CEMP stars at [Fe/H]$<-5$ are 
%imprinted by zero metallicity AGB stars. Thus, they have to show 
%the typical enhancement in s-process elements. This provides a precise prediction
%for the existence of  Pop III stars with $m_{\rm popIII} < 10 M_\odot$, that can be tested 
%by increasing the statistics of CEMP 
%stars with available s-process measurements at [Fe/H]$<-5$ \citep[e.g.][]{bonifacio15}.

Given the current statistics, we show that a flat Pop~III IMF with
$m_{\rm popIII} = [10 - 300]\, \Msun$ is disfavoured by the observations.
Furthermore, by assuming a Larson IMF with $m_{\rm popIII} = [0.1 - 300]\, \Msun$
and $m_{\rm ch} = 0.35 \, \Msun$, we find that on average $\approx 35 \%$ of 2G CEMP
stars at [Fe/H]$<-5$ are imprinted {\it also} by the chemical products of zero
metallicity AGB stars. Such Pop~III AGB stars can provide up to $30 \%$
of the metals polluting the ISM of formation of these CEMP stars, which therefore
might show the typical enhancement in s-process elements. This provides a prediction
for the existence of Pop~III stars with $m_{\rm popIII} < 10 M_\odot$, that can be
tested by increasing the statistics of CEMP stars with available s-process measurements
at [Fe/H]$<-5$ \citep[e.g.][]{bonifacio15}.

As a final point, we recall that, with our fiducial model, we find a larger number of C-normal stars 
with $-5<$[Fe/H]$<-4$ than observed.
As a consequence, in this [Fe/H] range the fraction of CEMP-to-C-normal 
stars is also lower than observed, although consistent with the data at 
$1-\sigma$ level (Fig.~\ref{fig:fiducial}). Several solutions for this small discrepancy 
do exist:\\
(i) the global contribution of Pop~III stars to metal-enrichment might
have been underestimated in the model, which do not account for the 
{\it inhomogeneous mixing of metals} into the MW environment. Including this
physical effect would natural delay the disappearance of Pop~III stars \citep{salvadori14};\\
(ii) a fraction of Pop~II stars with $m_{\rm popII} = [10 - 40]\, \Msun$ may evolve as faint 
SNe rather than normal core-collapse SNe, thus further contributing to enrich 
the gaseous environments with C, and reducing the formation of C-normal stars 
\citep{debennassuti14};\\ 
(iii) another solution concerns {\it chemical feedback}. If we exclude the 
Caffau's star, the low-Fe tail of C-normal stars is consistent with 
$Z_{\rm cr} \approx 10^{-4.5}\, \Zsun$, which means that these low-mass relics form
{\it thanks to dust} but in environments that might correspond to higher ${\cal D}_{\rm cr}$ 
(or lower $f_{\rm dep}$) than we assumed here, where $Z_{\rm cr}={\cal D}_{\rm cr}/f_{\rm dep}$
and $\cal{D}_{\rm cr}$ can be expressed as \citep{schneider12a}:
\begin{equation}
{\cal D}_{\rm cr} = [2.6 - 6.3] \times 10^{-9} \left[\frac{T}{10^3 \mbox{K}}\right]^{-1/2} \left[\frac{n_{\rm H}}{10^{12} \mbox{cm}^{-3}}\right]^{-1/2}.
\label{eq:dustcrit}
\end{equation}
\noindent
Here we have assumed the total grain cross section per unit mass of dust to vary in the range 
$2.22 \le S/10^5 \mbox{cm}^2/\mbox{gr} \le 5.37$, and a gas density
and temperature where dust cooling starts to be effective equal to $n_{\rm H} = 10^{12} \, {\rm cm}^{-3}$ and 
$T = 10^3\,$~K . Since $S$ could vary in a broader interval depending on the properties of the SN progenitor, 
this might lead to a larger variation of the value of ${\cal D}_{\rm cr}$;\\
%The latter values may well depend on
%$z$ and on the metallicity of the star-forming regions, and these variations might lead to up to one 
%order of  magnitude larger (lower) value of ${\cal D}_{\rm cr}$; \\
(iv) since C-normal stars at $-5 \le \rm [Fe/H] \le -4$ predominantly form in MW progenitors which have ${\cal D} \ge {\cal D}_{\rm cr}$ and have accreted 
their heavy elements from the MW environment (as opposed to haloes that have been self-enriched by previous stellar generations), we might have 
overestimated the ${\cal D}$ of {\it accreted material}. In fact, 
no destruction is assumed to take place in SN-driven outflows, when the 
grains are mixed in the external medium, or during the phase of accretion 
onto newly formed haloes. Indeed, the null detection of C-normal stars 
with $\rm [Fe/H] < -4.5$, beside the Caffau's star and despite extensive searches, might be an indication that,
at any given $Z$, haloes accreting their heavy 
elements from the MW environment might be less dusty than self-enriched haloes;\\
(v) a final possibility pertains the effect of {\it inhomogeneous radiative 
feedback}, which might reduce (enhance) the formation of C-normal (C-enhanced)
stars in mini-haloes locally exposed to a strong (low) LW/ionizing radiation. 
These effects are expected to be particularly important at high-$z$, i.e. before 
the formation of a global uniform background \citep{graziani15}.\\
Although all these solutions are plausible, we should not forget that our
comparison is actually based on $10$ stars at [Fe/H]$<-4$, which makes
the intrinsic observational errors larger than those induced by different
merger histories of the MW (Fig.~\ref{fig:fiducial}, upper panel). This underlines the quest
for more data to better understand the intricate network of physical processes 
driving early galaxy formation.

%%%%%%%%%%%%%%%%%%%%%%%%%%%%%%%%%%%%%%%%%%%%%%%%%%%%%%%%%%%%%%%%%%%%
\section*{Acknowledgments}
We acknowledge D. Yong for kindly sharing his data, P.~Bonifacio, 
T.~Beers, N.~Christlieb, and P.~Molaro for useful discussions, 
and the anonymous referee for a constructive and useful report. 
We thank the Kapteyn Institute of Groningen for hosting M. de 
Bennassuti during the development of this project and the Kavli 
Institute for Theoretical Physics for hosting K. Omukai, S. Salvadori 
and R. Schneider during the completion of the work. The research 
leading to these results has received funding from the European 
Research Council under the European Union’s Seventh Framework 
Programme (FP/2007-2013)/ERC Grant Agreement n. 306476, from the 
National Science Foundation under Grant No. NSF PHY11-25915, and 
from the JSPS KAKENHI Grant No. 25287040. S. Salvadori was partially 
supported by the Netherlands Organization for Scientific Research 
(NWO) through a VENI grant 639.041.233, and by the European Commission
through a Marie Sklodowska-Curie Fellowship, project PRIMORDIAL 700907.
%%%%%%% BIBLIOGRAPHY %%%%%%%%%%%%%%%%%%%%%%%%%%%%%%%%%%%%%%%%%%%%%%%%%
\bibliographystyle{mn2e}
\bibliography{salvadori}
\label{lastpage}

\end{document}